\documentclass[aps,showpacs,nofootinbib,amssymb,preprint]{revtex4}
\usepackage{graphicx,epsfig}
\usepackage{simplewick}
\usepackage{slashed}

\usepackage{booktabs}
\usepackage{latexsym}
\usepackage{amsmath}
\usepackage{amsfonts}
\usepackage{amssymb}
\usepackage{graphicx}
\newcommand{\nn}{\nonumber}
\newcommand{\be}{\begin{equation}}
\newcommand{\ee}{\end{equation}}
\newcommand{\bea}{\begin{eqnarray}}
\newcommand{\eea}{\end{eqnarray}}

\def\als{\alpha_{\rm s}}
\def\siml{{\ \lower-1.2pt\vbox{\hbox{\rlap{$<$}\lower6pt\vbox{\hbox{$\sim$}}}}\ }}

\newcommand{\MS}{\overline{\rm MS}}
\def\lQ{\Lambda_{\rm QCD}}
\newcommand{\RS}{\rm RS}

\newcommand{\ra}[1]{\renewcommand{\arraystretch}{#1}}

\begin{document}
\title{\boldmath 
Improved determination of heavy quarkonium magnetic dipole transitions in pNRQCD
\unboldmath}
\author{Antonio Pineda$^{(1)}$ and J. Segovia$^{(2)}$}
\affiliation{${}^{(1)}$Grup de F\'\i sica Te\`orica, Universitat
Aut\`onoma de Barcelona, E-08193 Bellaterra, Barcelona, Spain}
\affiliation{${}^{(2)}$Physics Division, Argonne National Laboratory, Argonne, Illinois 60439,
USA}

\date{\today}

\begin{abstract}
\noindent
We compute the magnetic dipole transitions between low-lying heavy quarkonium states in a 
model-independent way. We use the weak-coupling version of the effective field theory named potential NRQCD with the static 
potential exactly incorporated in the leading order Hamiltonian. 
The precision we reach is $k_{\gamma}^3/m^2\times{\cal O}(\als^2,v^2)$ and $k_{\gamma}^3/m^2\times{\cal O}(v^4)$ for the allowed and forbidden transitions respectively, where $k_{\gamma}$ is the photon energy. We also resum the large logarithms associated with the heavy quark mass scale. The specific transitions considered in this paper are the following: 
$\Upsilon(1S) \to \eta_b(1S)\,\gamma$,  $J/\psi(1S) \to \eta_c(1S) \,\gamma$, $h_b(1P) \to \chi_{b0,1}(1P)\,\gamma$, 
$\chi_{b2}(1P) \to h_b(1P) \,\gamma$, $\Upsilon(2S) \to \eta_b(2S)\,\gamma$, $\Upsilon(2S) \to \eta_b(1S)\,\gamma$ and 
$\eta_b(2S)\to\Upsilon(1S)\,\gamma$. 
The effect of the new power counting is found to be large and the exact treatment of the soft logarithms of the static potential makes the factorization scale dependence much smaller. The convergence for the $b\bar b$ ground state is quite good, and also quite reasonable for the $c\bar c$ ground state and the $b\bar b$ $1P$ state. For all of them we give solid predictions. For the $2S$ decays the situation is less conclusive, yet our results are perfectly consistent with existing data, as the previous disagreement with experiment for the $\Upsilon(2S) \to \eta_b(1S)\,\gamma$ decay fades away. We also compute some expectation values like the electromagnetic radius, $\langle r^2 \rangle$, or $
\langle { p}^2 \rangle$. We find $\langle r^2 \rangle$ to be nicely convergent in all cases, whereas the convergence of 
$\langle { p}^2 \rangle$ is typically worse.
%\vspace{2mm} \\
\end{abstract}
\maketitle

\section{Introduction}
\label{sec:Introduction}

Heavy quarkonium has always been thought to be the "hydrogen atom" of QCD. 
The reason is that the heavy quarks in the bound state move at nonrelativistic velocities: $v \ll 1$.
This allows testing the dynamics associated with the gluonic and light-quark degrees of freedom in a kinematic regime otherwise unreachable with only light degrees of freedom. 
Effective field theories (EFT's) directly derived from
QCD, like NRQCD~\cite{Caswell:1985ui} or pNRQCD~\cite{Pineda:1997bj}
(for some reviews see Refs.~\cite{Brambilla:2004jw,Pineda:2011dg}) disentangle the dynamics of the 
heavy quarks from the dynamics of the light degrees of freedom  efficiently and in a model-independent way.  
They profit from the fact that the dynamics of the bound state system is   
characterized by, at least, three widely separated scales: hard (the
mass $m$ of the heavy quarks), soft (the relative momentum $|{\vec p}|
\sim mv \ll m$ of the heavy-quark--antiquark pair in the center-of-mass frame), and ultrasoft (the typical kinetic energy $E \sim
mv^2$ of the heavy quark in the bound state system).

In this paper we use pNRQCD. This EFT takes full advantage of the
hierarchy of scales that appear in the system,
\begin{equation}
\label{hierarchy}
m \gg mv \gg mv^2 \cdots
\,,
\end{equation}
and makes a systematic and natural connection between quantum
field theory and the Schr\"odinger equation. Schematically the EFT
takes the form
\begin{eqnarray*}
%\vspace{-0.4in}
\,\left.
\begin{array}{ll}
&
\displaystyle{
\left(i\partial_0-{{\vec p}^2 \over m}-V_s^{(0)}(r)\right)\phi({\vec r})=0}
\\
&
\displaystyle{\ + \ \mbox{corrections to the potential}}
\\
&
\displaystyle{\ +\ 
\mbox{interactions with other low-energy degrees of freedom}}
\end{array} \right\}
{\rm pNRQCD}
\end{eqnarray*}
where $V_s^{(0)}(r)$ is the static potential and $\phi({\vec r})$ is
the $Q$-$\bar{Q}$ wave function.

The specific construction details of pNRQCD are slightly different depending on the relative size between the soft and the 
$\lQ$ scale. Two main situations are distinguished, namely, the weak-coupling \cite{Pineda:1997bj,Brambilla:1999xf} ($mv \gg \lQ$)  and the strong-coupling \cite{Brambilla:2000gk} ($mv \simeq \lQ$) versions of pNRQCD.
One major difference between them is that in the former the potential can be computed in perturbation theory unlike in the latter. 

It is obvious that the weak-coupling version of pNRQCD is amenable for a theoretically much cleaner analysis. The functional dependence on the parameters of QCD ($\als$ and the heavy quark masses) is fully under control and directly derived from QCD. The observables can be computed in well-defined expansion schemes with increasing accuracy, and nonperturbative effects are 
$\sim e^{-1/\als}$, exponentially suppressed compared with the expansion in powers of $\als$. Therefore, observables that could be computed with the weak-coupling version of 
pNRQCD are of the greatest interest. They may produce stringent tests of QCD in the weak-coupling regime (but yet with an all-order resummation of powers of $\als$ included) and, precision permitting, are ideal places in which to accurately determine some of the parameters of QCD. Nowadays there seems to be a growing consensus that the weak-coupling regime works properly for $t$-$\bar
t$ production near threshold, the bottomonium ground state mass, and bottomonium sum rules. To reach this conclusion, it is crucial to properly incorporate renormalon effects, which leads to convergent series, and the resummation of large logarithms, which significantly diminish the factorization scale dependence of the observable. Nevertheless, even in those cases, the situation is not optimal. For some 
observables, even if getting a convergent expansion, the corrections are large, or in the case of the bottomonium 
ground state hyperfine splitting a two-sigma level tension between experiment (see Ref.~\cite{Beringer:1900zz}) and theory \cite{Kniehl:2003ap,Recksiegel:2003fm} exists. 

In order to improve the convergence properties of the theory, the perturbative expansion in pNRQCD was 
rearranged in Ref. \cite{Kiyo:2010jm}. In this new expansion scheme the static potential was exactly included in the leading order (LO) Hamiltonian. The motivation behind this reorganization of the perturbative series is the observation \cite{Pineda:2002se} that, when comparing the static potential with lattice perturbation theory, one finds a nicely convergent sequence to the lattice data (at short distances). Yet, for low orders, the agreement is not good and the incorporation of corrections is compulsory to get a good agreement. This effect can be particularly important in observables that are more sensitive to the shape of the potential, and it naturally leads us to consider a double expansion in powers of $v$ and $\als(m)$, where $v$ has to do with the expectation value of the kinetic energy (or the static potential) in this new expansion scheme. 

In Ref. \cite{Kiyo:2010jm} this new expansion scheme 
was applied to the computation of the heavy quarkonium inclusive electromagnetic decay ratios. 
An improvement of the 
convergence of the sequence for the top and bottom cases was observed. It was particularly remarkable that the exact incorporation of the static potential allowed one to obtain agreement between theory and experiment for the case of the charmonium ground state. This leads to the second motivation of the present study: the possible applicability of the weak-coupling version of pNRQCD to the charmonium (ground state) and the $n=2$ 
excitation of the bottomonium. For those states the situation is more uncertain. Whereas Refs. \cite{Beneke:2005hg,GarciaiTormo:2005bs,DomenechGarret:2008vk} claimed that it is not possible to describe the bottomonium higher excitations  
in perturbation theory, an opposite stand is taken in Refs. \cite{Brambilla:2001fw,Brambilla:2001qk,Recksiegel:2002za,Recksiegel:2003fm}. We hope that we may shed some light on this issue as well.

The above discussion basically refers to the determination of the heavy quarkonium mass and inclusive electromagnetic decay widths. 
Obviously there are more observables that can be considered. Some of those are the radiative transitions: $H(n) \rightarrow H(n')\gamma$, 
where $n$, $n'$ stand for the principal quantum numbers of the heavy quarkonium. 
In Ref. \cite{Brambilla:2005zw} the allowed ($n=n'$) and hindered ($n\not= n'$) magnetic dipole (M1) transitions between low-lying heavy quarkonium states 
were studied with pNRQCD in the strict weak-coupling limit. The authors of that work also performed a 
detailed comparison of the EFT and potential model (see Refs.~\cite{Voloshin:2007dx,Eichten:2007qx}
for some reviews) results.
The specific transitions considered in that paper were the following: 
$J/\psi(1S) \to \eta_c(1S)\,\gamma$, 
$\Upsilon(1S) \to \eta_b(1S)\,\gamma$, $\Upsilon(2S) \to \eta_b(2S)\,\gamma$, $\Upsilon(2S) \to \eta_b(1S)\,\gamma$,
$\eta_b(2S)\to\Upsilon(1S)\,\gamma$, $h_b(1P) \to \chi_{b0,1}(1P)\,\gamma$ and $\chi_{b2}(1P) \to h_b(1P) \,\gamma$. Large errors were assigned to the pure ground state observables, especially for charmonium, whereas disagreement with experimental bounds (at that time) was found for the hindered transition $\Upsilon(2S) \to \eta_b(1S)\,\gamma$. 
In this paper we apply the new expansion scheme to those observables. 
The precisions we reach are $k_{\gamma}^3/m^2\times{\cal O}(\als^2,v^2)$ and $k_{\gamma}^3/m^2\times{\cal O}(v^4)$ for the allowed and forbidden transitions respectively, where $k_{\gamma}$ is the photon energy. Large hard logarithms (associated with the heavy quark mass) have also been 
resummed when they appear. 
The effect of the new power counting is found to be large and the exact treatment of the soft logarithms of the static potential makes the factorization scale dependence much smaller. 
The convergence for the $b\bar b$ ground state is quite good. This allows us to give a solid prediction for the $\Upsilon(1S) \to \eta_b(1S)\,\gamma$ transition with small errors. 
The convergence is also quite reasonable for the $c\bar c$ ground state and the $b\bar b$ $1P$ state. For all of them we give solid predictions. For the $J/\psi(1S)\to \eta_c(1S)\,\gamma$ transition our central value is significantly different from the one obtained in Ref. \cite{Brambilla:2005zw}, though perfectly compatible within errors. For the $2S$ decays the situation is less conclusive. Whereas for the $\Upsilon(2S) \to \eta_b(2S)\,\gamma$ decay we do not find convergence, previous disagreement with experiment for the hindered transition $\Upsilon(2S) \to \eta_b(1S)\,\gamma$ fades away with the new expansion scheme. 

The above observables depend on the expectation values of some quantum mechanical operators, like 
${\vec p}^{\,2}$ or $\vec{r}^{\,2}$ (the electromagnetic radius). Studying them in an isolated way is interesting on its own. First, they provide us with a very nice check of the renormalon dominance picture. According to this picture the determination of the heavy quarkonium mass using the static potential (in the on-shell scheme) should yield a bad convergent series, as is actually observed. The reason for this bad behavior is the existence of an $r$-independent constant that contributes to the potential and deteriorates the convergence of the perturbative series. If this is so, a check of this picture would be the computation of observables that are not affected by adding a constant to the potential. For those good convergence is expected. 
This is actually the case of  $\langle {\vec p}^{\,2} \rangle$ or $\langle {\vec r}^{\,2} \rangle$. We nicely see in Sec. \ref{sec:r2} that this picture is confirmed. 
We find the electromagnetic radius (somewhat surprisingly) to be nicely convergent in all cases. This allows us to talk of the typical (electromagnetic) radius of the bound state in those cases. The kinetic energy is also (though typically less than the radius) convergent
except for the $2S$ state. Then, we can also define a typical velocity $v\equiv \sqrt{\langle {\vec p}^{\,2} \rangle/m^2}$ for those states.

This paper is distributed as follows. In Sec. \ref{sec:th} we discuss the theoretical background of the computation and 
display the formulas we use for the decays. In Sec. \ref{sec:r2} we analyze $\langle {\vec p}^{\,2} \rangle$ and  $\langle {\vec r}^{\,2} \rangle$, 
and discuss renormalon dominance. In Sec. \ref{sec:decays} we compute the radiative transitions. Finally, in Sec. \ref{sec:con} we 
summarize our main results and give our conclusions.

\section{Theoretical setup}
\label{sec:th}
For the
purposes of this paper we can skip most details of pNRQCD. We will
only need the singlet static potential $V_s^{(0)}(r) \rightarrow V(r)$ and the spin-dependent
potential ${V}^{(2)}_{S^2,s}(r)\rightarrow V_{S^2}(r)$\footnote{For simplicity, we omit the index "$s$" for singlet and the 
upper indices "(0)" and "(2)" throughout the paper.}.
The static potential will be
treated exactly by including it in the leading-order Hamiltonian
\begin{eqnarray}
\label{H0}
H^{(0)}\equiv -\frac{{\bf \nabla}^2}{2m_r}+V(r), \qquad {\rm and} \qquad H^{(0)}\phi_{nl}({\vec r})=E_{nl}\phi_{nl}({\vec r})
\,,
\end{eqnarray}
where $m_r=m_1m_2/(m_1+m_2)$ (in this paper $m_1=m_2=m$). The static potential will be approximated by a polynomial of 
order $N+1$  in 
powers of $\als$ ($C_f=(N_c^2-1)/(2N_c)$, $C_A=N_c$)
\begin{eqnarray}
V^{(N)}(r)
&=&
 -\frac{C_f\,\alpha_s(\nu)}{r}\,
\bigg\{1+\sum_{n=1}^{N}
\left(\frac{\alpha_s(\nu)}{4\pi}\right)^n a_n(\nu;r)\bigg\}
\,.
\label{VSDfo}
\end{eqnarray}
In principle, we would like to take $N$ as large as possible (though we also want to explore the dependence on $N$). 
In practice, we take the static potential, at most, up to N=3, i.e., up to
${\cal O}(\als^4)$ including also the leading ultrasoft corrections.
This is the order to which the coefficients $a_n$ are completely known:
\begin{eqnarray}
a_1(\nu;r)
&=&
a_1+2\beta_0\,\ln\left(\nu e^{\gamma_E} r\right)
\,,
\nonumber\\
a_2(\nu;r)
&=&
a_2 + \frac{\pi^2}{3}\beta_0^{\,2}
+\left(\,4a_1\beta_0+2\beta_1 \right)\,\ln\left(\nu e^{\gamma_E} r\right)\,
+4\beta_0^{\,2}\,\ln^2\left(\nu e^{\gamma_E} r\right)\,
\,,
\nonumber \\
a_3(\nu;r)
&=&
a_3+ a_1\beta_0^{\,2} \pi^2+
\frac{5\pi^2}{6}\beta_0\beta_1 +16\zeta_3\beta_0^{\,3}
\nonumber \\
&+&\bigg(2\pi^2\beta_0^{\,3} 
+ 6a_2\beta_0+4a_1\beta_1+2\beta_2
+\frac{16}{3}C_A^{\,3}\pi^2\bigg)\,
  \ln\left(\nu e^{\gamma_E} r\right)\,
\nonumber \\
&+&\bigg(12a_1\beta_0^{\,2}+10\beta_0\beta_1\bigg)\,
  \ln^2\left(\nu e^{\gamma_E} r\right)\,
+8\beta_0^{\,3}  \ln^3\left(\nu e^{\gamma_E} r\right)\,
\nn
\\
&+&\delta a_3^{us}(\nu,\nu_{us}).
\label{eq:Vr}
\end{eqnarray}
The ${\cal O}(\als)$ term was computed in Ref. \cite{Fischler:1977yf}, the ${\cal O}(\als^2)$  in Ref. \cite{Schroder:1998vy}, 
the ${\cal O}(\als^3)$ logarithmic term in Refs. \cite{Brambilla:1999qa,Kniehl:1999ud}, the light-flavor finite piece in Ref. 
\cite{Smirnov:2008pn}, and the pure gluonic finite piece in Refs. \cite{Anzai:2009tm,Smirnov:2009fh}.  For the
ultrasoft corrections to the static potential we take
\begin{equation}
\delta a_3^{us}(\nu,\nu_{us})
%={8 C_A^3\over
%  3\beta_0}\pi^2 \ln\left(
%\alpha_{s}(\nu)\over \alpha_{s}(\nu_{us}) \right)
= \frac{16}{3}C_A^3 \pi^2\ln\left(\frac{\nu_{us}}{\nu}\right)
\, .
\end{equation}
We will not use the renormalization group improved ultrasoft
expression in this paper \cite{Pineda:2000gza,Hoang:2002yy,Eidemuller:1997bb,Brambilla:2009bi,Pineda:2011db}, 
as its numerical impact is small compared with other sources of error.

We will always work with three light (massless) quarks. For the case of the bottomonium ground state we also incorporate the leading 
effect due to the charm mass:
\begin{equation}
\delta V^{[2]}(r) = -\frac{4}{3}
\frac{\alpha_{s}^{(3)}(\nu)}{r}\left(\frac{\alpha_{s}^{(3)}(\nu)}{3\pi}
\right) \int_{1}^{\infty}dx\,
\frac{\sqrt{x^{2}-1}}{x^{2}}\left(1+\frac{1}{2x^{2}}
\right) e^{-2m_{c}rx}\,,
\label{eq:vpmc2}
\end{equation}
which can be easily read from the analogous QED computation (see, for instance, Ref. \cite{Pachucki:1996zza}). 
Its effect will be quite tiny. Therefore, we have only incorporated 
Eq. (\ref{eq:vpmc2}) in our final ($N=3$) evaluations and have not considered any other subdominant effects in the charm mass. 

The spin-dependent potential will be treated as a perturbation. It will contribute to the 
hindered M1 transitions. Nowadays, it is known with next-to-leading-log (NLL) accuracy \cite{Penin:2004xi}. 
Nevertheless, for consistency with our accuracy,  we will use its LL expression 
\begin{equation}
\label{Vspin}
V_{S^2}(\vec{r}\,) = \frac{4}{3} \pi C_{f} D_{S^{2},s}^{(2)}(\nu)
\delta^{(3)}(\vec{r}\,)
\end{equation}
where \cite{Pineda:2001ra} (see also \cite{Manohar:1999xd} for the derivation in vNRQCD)
\begin{equation}
D_{S^{2},s}^{(2)}(\nu) = \alpha_{s}(\nu)c_{F}^{2}(\nu)-\frac{3}{2\pi
C_{f}}\left(d_{sv}(\nu)+C_{f}d_{vv}(\nu)\right)
\end{equation}
depends on the NRQCD Wilson coefficients. With LL accuracy they read
\begin{equation}
\begin{split}
c_{F}(\nu) &= z^{-C_{A}}, \\
d_{sv}(\nu) &= d_{sv}(m), \\
d_{vv}(\nu) &=
d_{vv}(m)+\frac{C_{A}}{\beta_{0}-2C_{A}}\pi\alpha_{s}(m)(z^{\beta_{0}-2C_{A}}-1)
, 
\end{split}
\end{equation}
where
\begin{equation}
\begin{split}
z &=
\left[\frac{\alpha_{s}(\nu)}{\alpha_{s}(m)}\right]^{\frac{1}{\beta_{0}}}
\simeq 1-\frac{1}{2\pi}\alpha_{s}(\nu)\ln\left(\frac{\nu}{m}\right), \\
d_{sv}(m) &= C_{f}\left(C_{f}-\frac{C_{A}}{2}\right)\pi\alpha_{s}(m), \\
d_{vv}(m) &= -\left(C_{f}-\frac{C_{A}}{2}\right)\pi\alpha_{s}(m).
\end{split}
\end{equation}

\medskip

\begin{figure}[t!]
\makebox[-10.00cm]{\phantom b}
\put(0,5){\epsfxsize=8truecm \epsffile{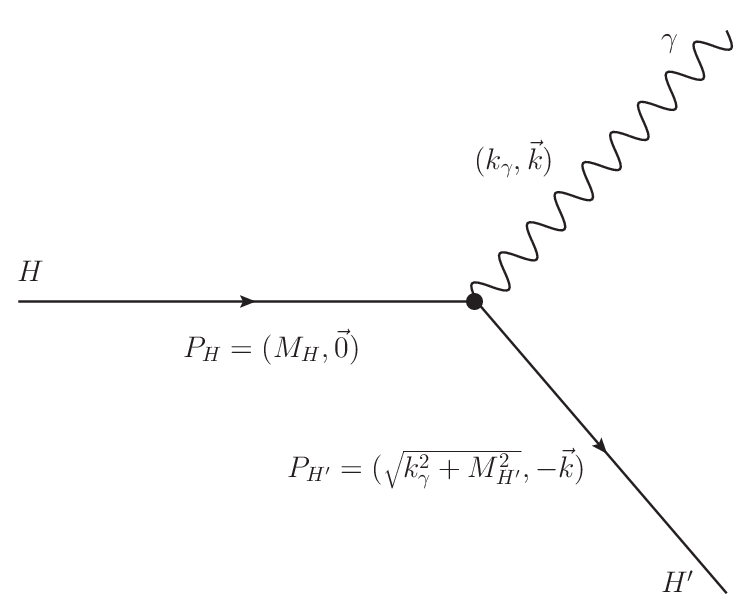}}
%\put(55,75){$P_{H} = (M_{H},\vec{0})$}
%\put(200,60){$P_{H'} = \left(\sqrt{k_{\gamma}^{2}+M_{H'}^{2}},-\vec{k}\right)$}
%\put(140,150){$(k_{\gamma},\vec{k})$}
\caption{\it
\label{fig:radiative} Kinematics of the radiative transition $H \to
H^\prime \gamma$ in the rest frame of the initial-state quarkonium $H$.}
\end{figure}

The theoretical study of the M1 transitions in the strict weak-coupling limit of 
pNRQCD has been carried out in detail in Ref. \cite{Brambilla:2005zw}. A 
particular relevant result was that nonperturbative effects, associated with the mixing with the octet field, were subleading and 
beyond present precision. We can use their results
in our power counting scheme with minor modifications (note that the dependence on the ultrasoft scale only enters marginally through the 
static potential). The expressions we use for the decays are the following (see Fig. \ref{fig:radiative} for the kinematics)\footnote{
In the following we use the notation 
$\left\langle\right.\!\!nS|\vec{p}^{\,2}|nS\!\!\left.\right\rangle=\langle {p}^{2} \rangle_{n0}$, 
$\left\langle\right.\!\!nP|\vec{p}^{\,2}|nP\!\!\left.\right\rangle=\langle {p}^{2} \rangle_{n1}$, 
$\left\langle\right.\!\!n'S|\vec{p}^{\,2}|nS\!\!\left.\right\rangle={}_{n'0}\langle {p}^{2} \rangle_{n0}$ 
and so on.}
\bea
\label{nS=nS}
\Gamma(n^{3}S_{1}\to n^{1}S_{0}\gamma) &=& \frac{4}{3}\alpha e_Q^{2}
\frac{k_{\gamma}^{3}}{m^{2}}
\left[(1+\kappa_{})^2 - \frac{5}{3} 
\frac{\langle {p}^{\,2} \rangle_{n0}}{m^{2}}
\right], \\
\label{nVnot=nP}
\Gamma(n^{3}S_{1}\to n'^{1}S_{0}\gamma) &\stackrel{n\neq n'}{=}&
\frac{4}{3}\alpha e_Q^{2} \frac{k_{\gamma}^{3}}{m^{2}} \left[
-\frac{k_{\gamma}^{2}}{24}{}_{n'0}\langle {r}^{2} \rangle_{n0}-\frac{5}{6}\frac{{}_{n'0}\langle {p}^{2} \rangle_{n0}}{m^
{2 } } +
\frac{2}{m^{2}}\frac{{}_{n'0}\langle V_{S^2}
(\vec{r}) \rangle_{n0}}{E_{n0}-E_{n'0}}\right]^{2}, \\
\label{nPnot=nV}
\Gamma(n^{1}S_{0}\to n'^{3}S_{1}\gamma) &\stackrel{n\neq n'}{=}& 4\alpha
e_Q^{2} \frac{k_{\gamma}^{3}}{m^{2}} \left[
-\frac{k_{\gamma}^{2}}{24}{}_{n'0}\langle {r}^{2} \rangle_{n0}-\frac{5}{6}\frac{{}_{n'0}\langle {p}^{2} \rangle_{n0}}{m^
{2 } } 
-
\frac{2}{m^{2}}\frac{{}_{n'0}\langle V_{S^2}
(\vec{r}) \rangle_{n0}}{E_{n0}-E_{n'0}}\right]^{2}, \\
\Gamma(n^{3}P_{J}\to n^{1}P_{1}\gamma) &
=&
\frac{3\Gamma(n^{1}P_{1}\to n^{3}P_{J}\gamma)}{2J+1}
= 
\frac{4}{3}\alpha e_Q^{2}
\frac{k_{\gamma}^{3}}{m^{2}}
\left[(1+\kappa_{})^2 - d_{J} 
\frac{\langle {p}^{2} \rangle_{n1}}{m^{2}}
\right], 
%\nn
%\\
\label{nP=nP}
\eea
where in Eq. (\ref{nP=nP}) $d_{0}=1$, $d_{1}=2$, $d_{2}=8/5$, 
\begin{equation}
k_{\gamma} = |\vec{k}| = \frac{M_{H}^{2}-M_{H'}^{2}}{2M_{H}}
\,,
\end{equation}
and the anomalous magnetic moment of the heavy quark, which is renormalization group invariant, reads
\be
\kappa_{}=\kappa^{(1)}\alpha_{s}(m)+\kappa^{(2)}\alpha^2_{s}(m)+\cdots
\ee
\bea
\kappa^{(1)} &=& 
\left(\frac{C_{f}}{2\pi}\right)
 \\
\kappa^{(2)} &=&\frac{C_{f}}{\pi^2} \left[\left(
-\frac{31}{16}+\frac{5\pi^{2}}{12}-\frac{\pi^{2}\ln 2}{2} + \frac{3\zeta_3}{4}
\right) C_{f} \right. \\
&&
\nn
\left.
 +
\left(\frac{317}{144}-\frac{\pi^{2}}{8}+\frac{\pi^{2}\ln
2}{4}-\frac{3\zeta_3}{8}\right) C_{A} +\left(-\frac{25n_{f}}{36}+\frac{119}{36}-\frac{\pi^{2}}{3}\right) T_{F}
\right].
\eea
We take $\kappa_{}$ from Ref. \cite{Grozin:2007fh} (it was originally computed in Refs. \cite{Fleischer:1992re,Bernreuther:2005gq}, though the first reference suffered from a factor 4 misprint).

Equations (\ref{nS=nS},\ref{nVnot=nP},\ref{nPnot=nV},\ref{nP=nP}) follow from the expressions obtained in Ref. \cite{Brambilla:2005zw}, except for 
the following changes: (i) The matrix elements of ${\vec r}^{\,2}$, ${\vec p}^{\,2}$ and $V_{S^2}$  are computed using the exact solution of Eq. (\ref{H0}) with the static potential approximated to the power $N$ instead of using the Coulomb potential; (ii) we use the heavy quark anomalous dimension $\kappa_{}$ to ${\cal O}(\als^2)$; (iii) our expression for $V_{S^2}$, Eq. (\ref{Vspin}), incorporates the LL resummation of logarithms (this will actually be important for the $2S \rightarrow 1S$ decays). Overall, our expressions are accurate with $k_{\gamma}^{3}/m^{2}\times{\cal O}(v^2,\als^2)$ and $k_{\gamma}^{3}/m^{2}\times{\cal O}(v^4)$ precision for the allowed and hindered transitions, respectively, and also include the resummation of large (hard) logarithms.  

Equations (\ref{nS=nS},\ref{nVnot=nP},\ref{nPnot=nV},\ref{nP=nP}) have been obtained in the on-shell scheme. Therefore, they depend on the pole mass $m$ and the static potential $V$, both of which suffer from severe renormalon ambiguities. On the other hand, the decays themselves are renormalon-free, as they are observables. Therefore, it is convenient to make the renormalon cancellation 
explicit. One first makes the substitution\footnote{Note that $\delta m_X$ and
$m$ (or $V$) have to be expanded to the {\it same} power in $\als$ and at the {\it same} scale.} 
\begin{equation}
\label{VsRen}
(m,V(r)) = (m_X+\delta m_X,V_{X}(r)-2\, \delta m_X)
\,,
\end{equation}
where 
\be
\delta m^{(N)}_X(\nu_f)=\nu_f\sum_{n=0}^N \delta m^{(n)}_X(\frac{\nu_f}{\nu}) \als^{n+1}(\nu)
\ee 
represents a residual mass that encodes the pole
mass renormalon contribution and $X$ stands for the specific
renormalon subtraction scheme.
Matrix elements are renormalon-free but not the heavy quark mass. Its renormalon ambiguity cancels with the one coming from the anomalous magnetic moment of the heavy quark.
The renormalon structure of the chromomagnetic moment of the heavy quark has been studied in detail in Ref. \cite{Grozin:1997ih}. If one does the Abelian-like limit, one can get the renormalon structure of the anomalous magnetic moment of the heavy quark. One sees that it suffers from the very same renormalon as the heavy quark mass. Therefore, the quantity $(1+\kappa_{})/m$ is free of the renormalon ambiguity (or at least of the leading one). When rewriting the decay expressions from the on-shell scheme to the $X$ scheme, the change is absorbed in $\kappa_{}$ so $\kappa_{} \rightarrow \kappa_{X}$ where
\be
\kappa_{X}=\kappa^{(1)}_{X}\alpha_{s}(m)+\kappa^{(2)}_{X}\alpha^2_{s}(m)+\cdots
\ee
with
\be
\kappa^{(1)}_{X}=\kappa^{(1)}-\frac{\nu_f}{m}\delta m^{(0)}_X(\frac{\nu_f}{m})
\ee
\be
\kappa^{(2)}_{X}=\kappa^{(2)}-\frac{\nu_f}{m}\delta m^{(1)}_X(\frac{\nu_f}{m})-\kappa^{(0)}\frac{\nu_f}{m}\delta m^{(0)}_X(\frac{\nu_f}{m})
+\left(\frac{\nu_f}{m}\delta m^{(0)}_X(\frac{\nu_f}{m})\right)^2
\,.
\ee
Overall in Eqs. (\ref{nS=nS},\ref{nVnot=nP},\ref{nPnot=nV},\ref{nP=nP}) we have to make the replacement $(m,V,\kappa_{})
\rightarrow (m_X,V_X,\kappa_{X})$ throughout. Note that, once written in terms of renormalon-free quantities, one may consider different $N$, $N'$ for $V_X^{(N)}$, $m_X^{(N')}$, ..., and the observable would still be renormalon-free.

\section{Applicability of weak coupling to heavy quarkonium}
\label{sec:r2}

The allowed M1 radiative transitions depend on $\langle { p}^2 \rangle_{nl}$ and, at higher orders, on other expectation values such as 
$\langle r^2 \rangle_{nl}$. Studying them gives us a hint of the applicability of the weak-coupling version of pNRQCD to those states, and a very nice check of the renormalon dominance picture. In this section we compute the bound state energy ($E_{nl}$), $\langle { p}^2 \rangle_{nl}$ and $\langle r^2 \rangle_{nl}$ for the charmonium ground state and for $n=1,2$ bottomonium states. In the cases where 
we have good convergence, we will be able to obtain well-defined values for $v_{nl}\equiv \sqrt{\langle p^2 \rangle_{nl}/m^2}$ and $\langle r^2 \rangle_{nl}$.

Our reference values for the charm and bottom masses are $m_b(m_b)=4.19$ \cite{Pineda:2006gx} and $m_c(m_c)=1.25$ \cite{Signer:2008da}, which we then transform to renormalon 
subtracted schemes like the RS, RS' \cite{Pineda:2001zq} or PS \cite{Beneke:1998rk}. We will mainly use the RS' scheme and leave the RS 
and PS schemes for partial checking (in 
particular that the dependence on the renormalon subtraction scheme is small). Therefore, we use ($\delta m^{(0)}_{\RS'}=0$ and 
$d_n(\nu,\nu_f)=\beta_n/2^{1+2n}\ln\left(\frac{\nu}{\nu_f}\right)$)
\begin{equation}
\begin{split}
&
\delta m^{(1)}_{\RS'}(\frac{\nu_f}{\nu})= N_m
\frac{\beta_{0}}{2\pi} S(1,b), \\
&
\delta m^{(2)}_{\RS'}(\frac{\nu_f}{\nu})= N_m\left(\frac{\beta_{0}}{2\pi}\right)
\left[ S(1,b)
\frac{2d_{0}(\nu,\nu_{f})}{\pi} + \left(\frac{\beta_{0}}{2\pi}\right) S(2,b)
\right], \\
&
\begin{split}
\delta m^{(3)}_{\RS'}(\frac{\nu_f}{\nu}) =& N_m \left(\frac{\beta_{0}}{2\pi}\right)\times \\
&
\times \left[ 
S(1,b) \frac{3d_{0}^{2}(\nu,\nu_{f})+2d_{1}(\nu,\nu_{f})}{\pi^{2}} +
\left(\frac{\beta_{0}}{2\pi}\right) S(2,b) \frac{3d_{0}(\nu,\nu_{f})}{\pi} +
\left(\frac{\beta_{0}}{2\pi}\right)^{2} S(3,b) \right].
\end{split}
\end{split}
\end{equation}
where
\be
S(n,b)=\sum_{k=0}^2c_k\frac{\Gamma(n+1+b-k)}{\Gamma(1+b-k)}
\ee
with $c_0=1$ and 
\be
b={\beta_1 \over 2\beta_0^2}\,,
\qquad
c_1={1 \over 4\,b\beta_0^3}\left({\beta_1^2 \over \beta_0}-\beta_2\right)
\ee
and 
\be
c_2={1 \over b(b-1)}
{\beta_1^4 + 4 \beta_0^3 \beta_1 \beta_2 - 2 \beta_0 \beta_1^2 \beta_2 + 
   \beta_0^2 (-2 \beta_1^3 + \beta_2^2) - 2 \beta_0^4 \beta_3 
\over 32 \beta_0^8}
\,.
\ee
For easy of reference, we give some typical values that we will use in this paper: 
$m_{b,\RS'}(0.7\,{\rm GeV})=4902$ MeV, $m_{b,\RS'}(1\,{\rm GeV})=4859$ MeV, $m_{c,\RS'}(0.7\,{\rm GeV})=1648$ MeV, $m_{c,\RS'}(1\,{\rm GeV})=1536$ MeV. Our reference value for $N_m$ will be $N_m=0.574974$ (for three light flavors) from Ref. \cite{Pineda:2001zq}. To this number we will typically assign a 10\% uncertainty.
Our reference value for $\als$ will be $\als^{(n_f=3)}(1\; {\rm GeV})= 0.479778$, which we obtain running down $\als^{(n_f=5)}(M_Z)=0.118$. We then run with four loop accuracy for the typical scales of the bound state system. Unless stated otherwise, throughout the paper 
we will set $\nu_{us}=\nu_f$.

The static potential we will consider in the following will be (in the RS' scheme)
\begin{equation}
%\vspace{-0.4in}
\label{VRS}
V^{(N)}_{\RS'}(r)=
\,\left\{
\begin{array}{ll}
&
\displaystyle{
(V^{(N)}+2\delta m^{(N)}_{\RS'})|_{\nu=\nu}\equiv
 \sum_{n=0}^{N}V_{RS',n}\als^{n+1}(\nu)
\qquad {\rm if} \quad r>\nu_r^{-1} }
\\
&
\displaystyle{
(V^{(N)}+2\delta m^{(N)}_{\RS'})|_{\nu=1/r}\equiv
 \sum_{n=0}^{N}V_{\RS',n}\als^{n+1}(1/r)
\qquad {\rm if} \quad r<\nu_r^{-1}. }
\end{array} \right.
\end{equation}
This expression encodes all the possible limits:

{\bf (a)}.
 The case $\nu_r=\infty$, $\nu_f=0$ is nothing but the on-shell static potential at fixed order, i.e. Eq. (\ref{VSDfo}). 
 Note that the $N=0$ case reduces to a standard computation with a Coulomb potential, for which we can compare with analytic results for the matrix elements. We will use this fact to check our numerical solutions of the Schroedinger equation. 
If we also switch off the hard logs and the ${\cal O}(\als^2)$ correction to the anomalous magnetic moment, our computation would be equal to the one performed in Ref. \cite{Brambilla:2005zw}. We will use this fact to compare with their results throughout the paper. 

\begin{figure}[!t]
\begin{center}
\hspace{0.50cm}
\epsfig{figure=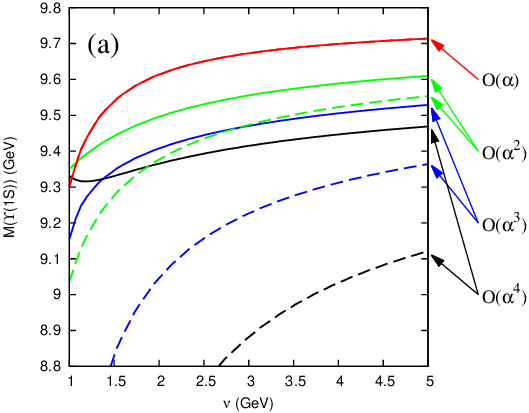,height=8.85cm,width=11.50cm} \\[3ex]
\epsfig{figure=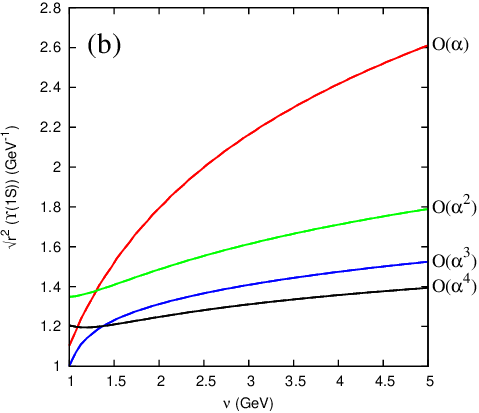,height=8.85cm,width=11.00cm}
\caption{\label{fig:botn1} \it Plot of $M_{10}=2 m_{b,\RS'}(0.7\,{\rm GeV})+E_{10}$  and $\sqrt{\langle r^2 \rangle_{10}}$ of the 
bottomonium ground state using the static potential $V_{\RS'}^{(N)}$ at different orders in perturbation theory: $N=0,1,2,3$. 
The dashed lines have been computed with $\nu_f=0$. The continuous lines have been computed with $\nu_f=0.7$ GeV.
In both cases $\nu_r=\infty$ GeV.}
\end{center}
\end{figure}

\begin{figure}[!t]
\begin{center}
\epsfig{figure=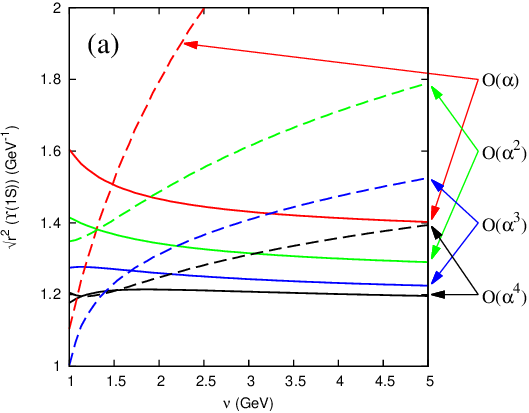,height=8.80cm,width=11.50cm} \\[3ex]
\epsfig{figure=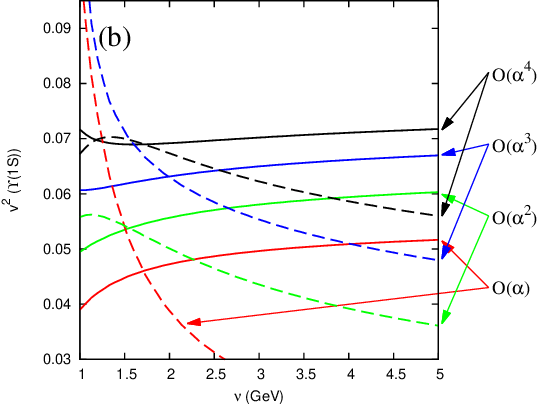,height=8.80cm,width=11.90cm}
\caption{\label{fig:botn11} \it Plot of $\sqrt{\langle r^2 \rangle_{10}}$  and $ v^2_{10}$ for the case of the 
bottomonium ground state using the static potential $V_{\RS'}^{(N)}$ at different orders in perturbation theory: $N=0,1,2,3$. 
The dashed lines have been computed with $\nu_r=\infty$. The continuous lines have been computed with $\nu_r=0.7$ GeV.
In both cases $\nu_f=0.7$ GeV.}
\end{center}
\end{figure}

{\bf (b)}.
The case $\nu_r=\infty$ (with finite non-zero $\nu_f$) is nothing but adding an r-independent constant to the static potential (see the discussion in Ref. \cite{Pineda:2002se}). Therefore, the results for $\langle {p}^2 \rangle_{nl}$ and $\langle r^2 \rangle_{nl}$ do not depend on the specific value of $\nu_f$ (for a fixed heavy quark mass). In particular, the value $\nu_f=0$ can be taken, which is equivalent to not considering any renormalon subtraction at all.
On the other hand, the binding energy $E_{nl}$ is renormalon dependent. This effect can be seen in full glory in Fig. \ref{fig:botn1}, where we plot  $M_{10}=2 m_{b,\RS'}(0.7\,{\rm GeV})+E_{10}$  and $\langle r^2 \rangle_{10}$ for the case of the 
bottomonium using the static potential $V_{\RS'}^{(N)}$ at different orders in perturbation theory: $N=0,1,2,3$. We clearly observe how, for the $\nu_r=\infty$, $\nu_f=0$ case, the bound state energy is not convergent (see dashed lines), whereas $\langle r^2 \rangle_{10}$ is (see solid lines). 
On the other hand, for the $\nu_r=\infty$, $\nu_f=0.7$ GeV case, both the bound state energy and $\langle r^2 \rangle_{10}$ show a nice 
convergent pattern as we increase $N$ (see solid lines). Note that $\langle r^2 \rangle_{10}$ is exactly the same in both cases: $\nu_f=0$ or $\nu_f=0.7$ (this is the reason only solid lines show up in Fig. 
\ref{fig:botn1}.b). The same analysis can be done for $\langle {p}^2 \rangle_{10}$, as one can see in 
Fig. \ref{fig:botn11} for the dashed lines (note, though, that $\langle {p}^2 \rangle_{10}$ is less convergent than $\langle {r}^2 \rangle_{10}$). 

\begin{figure}
\epsfig{figure=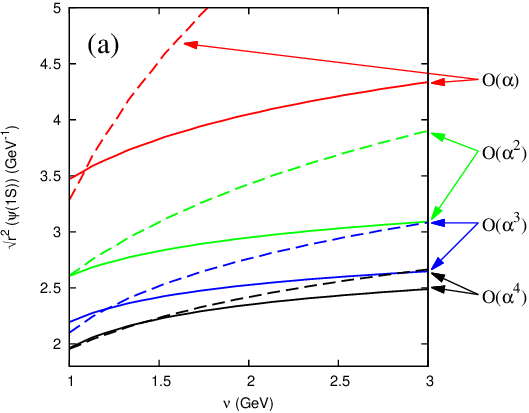,height=8.95cm,width=11.50cm} \\[3ex]
\epsfig{figure=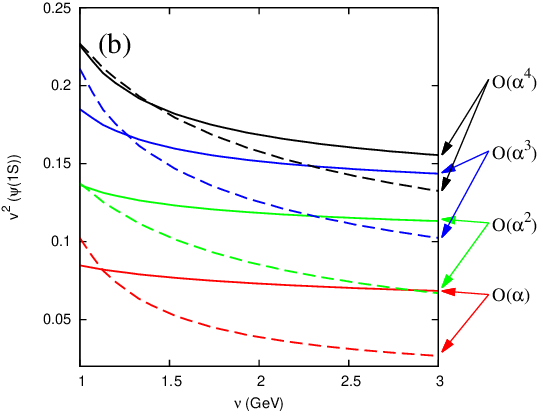,height=8.95cm,width=11.90cm}
\caption{\label{fig:charmn1} \it Plot of $\sqrt{\langle r^2 \rangle_{10}}$  and $v^2_{10}$ for the case of the 
charmonium using the static potential $V_{\RS'}^{(N)}$ at different orders in perturbation theory: $N=0,1,2,3$. 
The dashed lines have been computed with $\nu_r=\infty$. The continuous lines have been computed with $\nu_r=0.7$ GeV.
In both cases $\nu_f=0.7$ GeV.}
\end{figure}

A rather similar picture is observed for the charmonium ground state, though the sequences, as expected, 
are less convergent. Again,  
it is compulsory to incorporate the renormalon cancellation (finite $\nu_f$) to transform the bound state energy in a convergent sequence in $N$, whereas $\langle r^2 \rangle$ and $\langle {p}^2 \rangle$ are always convergent (see the dashed lines of Fig. \ref{fig:charmn1}); everything in full accordance with the renormalon dominance picture. 

\begin{figure}
\epsfig{figure=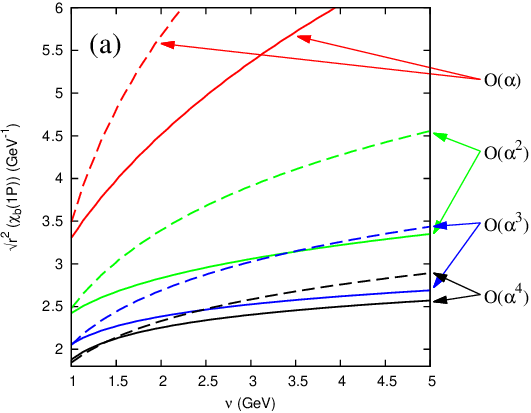,height=8.95cm,width=11.50cm} \\[3ex]
\epsfig{figure=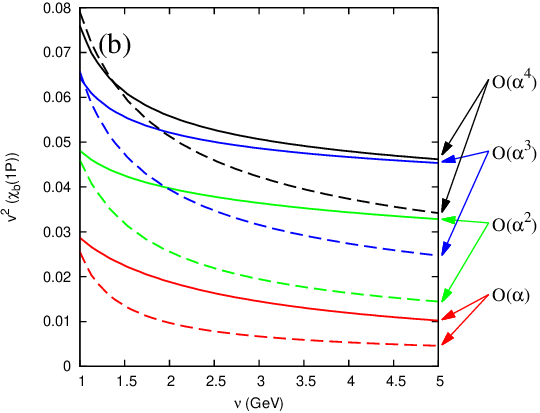,height=8.95cm,width=11.90cm}
\caption{\label{fig:botn21} \it Plot of $\sqrt{\langle r^2 \rangle_{21}}$  and $v^2_{21}$ for the case of the 
$n=2$ $P$-wave bottomonium using the static potential $V_{\RS'}^{(N)}$ at different orders in perturbation theory: $N=0,1,2,3$. 
The dashed lines have been computed with $\nu_r=\infty$. The continuous lines have been computed with $\nu_r=0.7$ GeV.
In both cases $\nu_f=0.7$ GeV.}
\end{figure}

\begin{figure}
\hspace{0.90cm}
\epsfig{figure=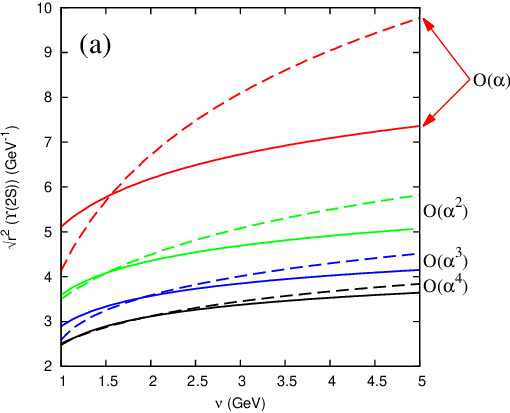,height=8.95cm,width=12.60cm} \\[3ex]
\epsfig{figure=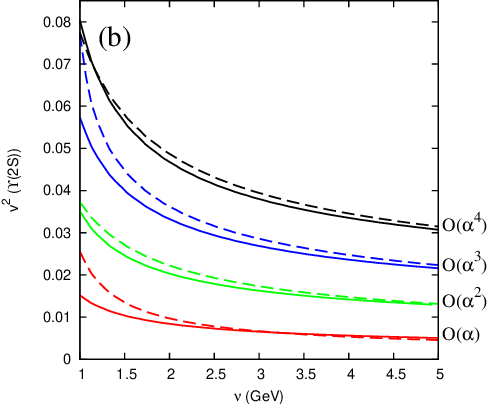,height=8.95cm,width=11.90cm}
\caption{\label{fig:botn20} \it Plot of $\sqrt{\langle r^2 \rangle_{20}}$  and $v^2_{20}$ for the case of the 
$n=2$ $S$-wave bottomonium using the static potential $V_{\RS'}^{(N)}$ at different orders in perturbation theory: $N=0,1,2,3$. 
The dashed lines have been computed with $\nu_r=\infty$. The continuous lines have been computed with $\nu_r=0.7$ GeV.
In both cases $\nu_f=0.7$ GeV.}
\end{figure}

For the $n=2$ bottomonium states the situation is less conclusive. For both the $P$- and $S$-wave $\langle r^2 \rangle$ is convergent, see 
 the dashed lines of Figs. \ref{fig:botn21}.a and \ref{fig:botn20}.a, respectively. On the other hand, $\langle {p}^2 \rangle$ is only marginally convergent for $P$-wave 
 (see the dashed lines of Fig. \ref{fig:botn21}.b), 
 or even not convergent for the $S$-wave (see the dashed lines of Fig. \ref{fig:botn20}.b), as each order is typically of the same size.

{\bf (c)}. We now take $\nu_r=finite$ (and, for consistency, $\nu_r \geq \nu_f$). We expect this case to improve over the previous results, as it incorporates the correct (logarithmically modulated) short distance behavior of the potential. Yet, 
this has to be done with care in order not to spoil the 
renormalon cancellation. For this it is compulsory from now on to keep a finite, nonvanishing, $\nu_f$;  otherwise, the renormalon cancellation is not achieved order by order in $N$, as was discussed in detail in 
Ref. \cite{Pineda:2002se}. We have explored the effect of different values of $\nu_f$ in our analysis. Large values of $\nu_f$ imply a large infrared cutoff. This makes our scheme to become closer to a $\MS$-like scheme. Such schemes still achieve renormalon cancellation, yet they jeopardize the power counting, as the residual mass does not count as $mv^2$. This comes at the cost of making the consecutive terms of the perturbative series bigger. Therefore, we prefer values of $\nu_f$ as low as possible, with the constraint that one should still obtain the renormalon cancellation, and that it is still 
possible to perform the expansion in powers of $\als$. 
In our analysis we observe that we can use a rather low value of $\nu_f$ and yet obtain the renormalon cancellation. By also taking a low value of $\nu_r$ we find that the convergence is accelerated and the scale dependence is significantly reduced. We illustrate this behavior in Figs. \ref{fig:botn11}, \ref{fig:charmn1}, \ref{fig:botn21}, and 
 \ref{fig:botn20}, where we can compare the case with $\nu_r=finite$ (continuous lines) and  $\nu_r=\infty$ (dashed lines).  
 This improvement is observed in all cases except for the $2S$ bottomonium $\langle {p}^2 \rangle_{20}$. Especially relevant 
 for us is that 
 it accelerates the convergence of $\langle {p}^2 \rangle_{10}$ for charmonium, and transforms $\langle {p}^2 \rangle_{21}$ into a convergent series.

Leaving aside the $2S$ bottomonium state, it is particularly appealing to compare the $N=0$ case for $\nu_r=finite$ and $\nu_r=\infty$. The latter corresponds to the Coulomb approximation, and it is the one used in the strict weak-coupling analysis performed for the radiative transitions in Ref. \cite{Brambilla:2005zw}. One can see a very strong scale dependence, almost a vertical line compared with the $\nu_r=0.7$ GeV case (see, for instance, Fig. \ref{fig:botn11}). Therefore, small variations of the scale produce very large changes in the theoretical prediction. This makes it difficult to assign central values (and errors). 
This is not the case after resumming the soft logarithms by setting $\nu_r\not=0$. This produces flatter plots. Note also that, typically, there is a scale where $\nu_r=\infty$ and $\nu_r=0.7$ lines cross. One can then take this scale 
 as a way to fix the scale $\nu$ of the computation with $\nu_r=\infty$ (which corresponds to the strict weak-coupling expansion).

Overall, we find the electromagnetic radius (somewhat surprisingly) to be nicely convergent in all cases. This allows us to talk of the typical (electromagnetic) radius of these bound states. The kinetic energy is also (though typically less than the radius) convergent,
except for the $2S$ state. Then, we can also define a $v^2_{nl}\equiv \langle p^2 \rangle_{nl}/m^2 $ for those states. 
We show these numbers in Table \ref{tab:v2}.  
These numbers can be taken as estimates of the typical radius of the bound state system and of the typical velocity of the 
heavy quarks inside the bound state. It is comforting that the numbers we obtain for $v^2$ are similar to those usually assigned either by potential models or by NRQCD (see, for instance, \cite{Quigg:1979vr,Bodwin:2007fz}). The specific values in the table have been
taken from the $N=3$ case at $\nu=1.5$ GeV (and $\nu_f=\nu_r=0.7$ 
GeV).  For the $b\bar b$ ground state the result is very stable under scale variations; for the 
charm ground state and for the bottomonium $P$-wave the scale dependence is bigger. We stress that the numbers in the table should be taken 
as estimates. We do not attempt here to perform a specific error analysis of those numbers, as it is not needed for the decays. Let us just mention that one source of the error would come from the $1/m$ subleading potentials. In principle, these effects would produce ${\cal O}(v^2)$ corrections. For bottomonium and charmonium this would typically mean
$\sim 7\%$ and $\sim 20\%$ variations of the central values, respectively. Especially for bottomonium, such uncertainties would compite with the difference between different $N$ evaluations or, in some cases, with the scale variation. Finally, we remark that for the $2S$ bottomonium state the numbers in the table should be taken 
with more caution, as there is no convergence in the sequence in $N$. 
One might actually be surprised by the fact that the $(n,l)=(2,1)$ and the $(n,l)=(2,0)$ states show this different behavior, as far as convergence is concerned, since the typical transfer momentum is the same. One should note, though, that the $(n,l)=(2,0)$ squared wave function has two maxima, and the most important one is a very low momentum. On the other hand, this problem only appears for  $\langle { p}^2 \rangle$ and not for $\langle r^2 \rangle$, so we find the situation inconclusive.

\begin{table}[h]
\ra{1.1}
\begin{center}
$$
\begin{array}{|l|c|c|c|c|}
\hline
   &b\bar b(1S)&c\bar c(1S)&b\bar b(1P)&b\bar b(2S)\\
\hline
v & 0.26 & 0.43 & 0.25 & 0.24 \\
\hline
\sqrt{\langle r^2 \rangle} ({\rm GeV^{-1}})& 1.2 & 2.2 & 2.1 & 2.9\\
\hline
\bottomrule
\end{array}
$$ 
\end{center}
\caption{\label{tab:v2}\it Estimates for $v \equiv \sqrt{\langle {p}^2 \rangle/m^2}$ and $\sqrt{\langle r^2 \rangle}$ for the heavy quarkonium states. For the $b\bar b(2S)$ state the number we give for $v$ is quite uncertain.}
\end{table}

The results of this and the following section have been obtained by solving the Schroedinger equation numerically. We have performed a series of tests of the numerical solutions. As we have already mentioned, the case $N=0$ with $\nu_r=\infty$ corresponds to the Coulomb case. We have checked the numerical solution against the known analytical result in this case. For a general $N$ and $\nu_r$ we have also computed $\langle p^2 \rangle$ either directly (in momentum space) or through the equality $\langle (E-V(r))\rangle=\langle  p^2/m\rangle$. Finally, we 
have also checked the wave function at the origin, either by direct computation (taking the smallest point at which the wave function has been 
computed and checking for stability) or through the equality $|\phi_{nl}(0)|^2=m/(4\pi)\langle V'(r)\rangle_{nl}$ 
(see Ref. \cite{Quigg:1979vr}).
 
\section{M1 Transitions}
\label{sec:decays}

In this section we compute the M1 radiative transitions for the low lying bottomonium and charmonium states. 

\subsection{$\Upsilon(1S) \rightarrow \eta_b(1S)\gamma$}
\label{sec:botn11}

Our central value for $\Gamma_{\Upsilon(1S) \rightarrow \eta_b(1S)\gamma}$ is obtained using Eq. (\ref{nS=nS}) with $N=3$, $\nu=1.5$ GeV, and $\nu_f=\nu_r=0.7$ GeV. For $k_\gamma$ we take the values of the $\Upsilon(1S)$ and $\eta_b(1S)$ 
masses from the PDG \cite{Beringer:1900zz}\footnote{We note, though, that there is a recent 
determination of the $\eta_b(1S)$ mass which is around 
10 MeV lower \cite{Mizuk:2012pb} than the PDG value. If such a value is confirmed 
$k_{\gamma}$ should be changed accordingly (as $\Gamma_{\Upsilon(1S) \rightarrow \eta_b(1S)\gamma} \propto k_{\gamma}^3$  the effect is important), which can be trivially done.}.
In table \ref{tab:Gammab1S} we show the size of the different contributions to $\Gamma_{\Upsilon(1S) \rightarrow \eta_b(1S)\gamma}$. 
The ${\cal O}(\als)$ and ${\cal O}(\als^2)$ corrections are evaluated at the mass scale. 
The ${\cal O}(\als(m))$ corrections in the RS' and on-shell scheme are equal. Renormalon effects first appear at 
${\cal O}(\als^2(m))$ and make this expansion more convergent. Yet, as we have taken a small value of $\nu_f$, the ${\cal O}(\als^2(m))$ 
term is still large. There are no ${\cal O}(v)$ corrections. 
The ${\cal O}(v^2)$ correction can be evaluated 
at different orders in $N$, and for different  values of the factorization scale. One can easily deduce its size by multiplying Fig. \ref{fig:botn11}.b by -5/3 times the LO result. The value quoted in Table \ref{tab:Gammab1S} for the 
${\cal O}(v^2)$ correction has been obtained for $N=3$ and $\nu=1.5$ GeV. An almost identical value is obtained if one takes 
$\nu$ to be the scale 
of minimal sensitivity. Actually, one also obtains a quite similar value if one takes the scale of minimal sensitivity of the $N=3$, $\nu_r=\infty$ computation. The great advantage of using $\nu_r=0.7$ GeV versus $\nu_r=\infty$ is that the $\nu$ dependence becomes very mild; thus, it is not a source of uncertainty, and one can give sensible predictions for the central values. We note that, depending on the order $N$, minimal sensitivity scales may not show up, as we can see in Fig. \ref{fig:botn11} for other values of $N$ and/or $\nu_r$. Therefore, in some cases such a prescription may not give a meaningful result and the series still be convergent. 
\begin{table}[hb]
\ra{1.1}
\begin{center}
$$
\begin{array}{|l|c|c|c|c|c|c|}
\hline
   &{\rm LO}&{\cal O}(\als)&{\cal O}(\als^2)&{\cal O}(v^2)&\als \times {\cal O}(\als^2)&v\times{\cal O}(v^2)\\
\hline
\delta \Gamma \;({\rm eV})& 14.87 & 1.29 & 0.73 & -1.71& 0.15 &-0.45\\
\hline
\bottomrule
\end{array}
$$ 
\end{center}
\caption{\label{tab:Gammab1S} \it The leading and subleading contributions to $\Gamma_{\Upsilon(1S) \rightarrow \eta_b(1S)\gamma}$. 
The last two numbers are error estimates obtained by multiplying the subleading ${\cal O}(\als^2)$ contribution by $\als$ 
and the subleading ${\cal O}(v^2)$ contribution by $v$.}
\end{table}

\begin{figure}
\includegraphics[width=0.7\textwidth]{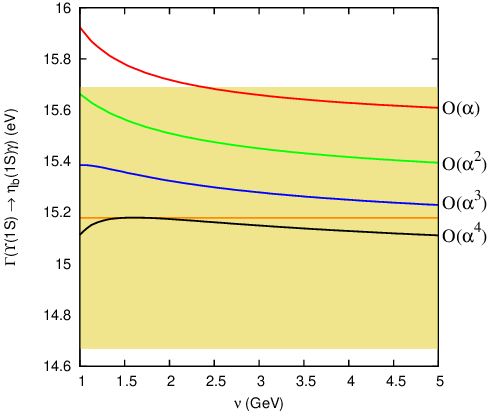}
\caption{\label{fig:botn1decay} \it Plot of $\Gamma_{\Upsilon(1S) \rightarrow \eta_b(1S)\gamma}$ for the case of the 
bottomonium ground state using the static potential $V_{\RS'}^{(N)}$ at different orders in perturbation theory: $N=0,1,2,3$ with $\nu_r=\nu_f=0.7$ GeV.
The horizontal line is our central value and the yellow band our final error estimate. 
}
\end{figure}

If we sum all the contributions of Table \ref{tab:Gammab1S} we obtain 15.18 eV, which is quite close to the LO 14.87 eV value. This is due 
to the strong cancellation between the
 $\als$ and $v$ corrections. The main source of uncertainty comes from higher order terms. 
 Because of the strong cancellation between the
 $\als$ and $v$ terms, we feel that adding a power of 
 $v$ to the overall correction would underestimate the error. Instead, we take the $v\times{\cal O}(v^2)$ contribution in Table \ref{tab:Gammab1S} as our estimate of the subleading correction, as it is the biggest possible contribution. Such term alone produces an error of order 3\%. This error is much bigger than the error one would obtain only considering scale variations (see Fig. \ref{fig:botn1decay}), or if we do the evaluation with $N=2$ instead of $N=3$ (see, again, Fig. \ref{fig:botn1decay}), or than the error associated with variations of $\nu_f$. All these errors are associated with higher order effects. 
We do not include those, in order to avoid double counting. The only other source of theoretical error that we include is the one due to $N_m$
(for the evaluation of this error we also take into account the correlation with the bottom mass value). Besides the theoretical 
error, we also include the error associated with the QCD parameters, even though its size is typically smaller than the theoretical error. For $\als$ we take the variation $\als(M_z)=0.118\pm0.001$ \cite{Beringer:1900zz}. For the variation of the $\MS$ bottom mass we take $m_b(m_b)=4.19\pm 0.03$ GeV. In summary, we obtain the following result for the different error contributions:
\be
\Gamma_{\Upsilon(1S) \rightarrow \eta_b(1S)\gamma}
=15.18 \pm 0.45({\cal O}(v^3)){}^{-0.12}_{-0.05}(N_m){}^{-0.04}_{+0.03}(\als){}^{-0.20}_{+0.20}(m_{\MS})  \;{\rm eV}
\,,
\ee
which after combining the errors in quadrature reads
\be
\label{Gamma1Setab}
\Gamma_{\Upsilon(1S) \rightarrow \eta_b(1S)\gamma}
=15.18(51) \;{\rm eV}
\,.
\ee
This corresponds to a branching fraction of $2.9\times10^{-4}$. Equation (\ref{Gamma1Setab}) is bigger than the result obtained in Ref. \cite{Brambilla:2005zw} ($\sim (k_{\gamma}/39)^3\times 2.5$ keV, see Ref. \cite{Vairo:2006pc}) but compatible within errors.

\subsection{$J/\psi(1S) \rightarrow \eta_{c}(1S)\gamma$}

Our central value for $\Gamma_{J/\psi(1S) \rightarrow \eta_{c}(1S)\gamma}$ is obtained using Eq. (\ref{nS=nS}) with $N=3$, $\nu=1.5$ GeV, and $\nu_f=\nu_r=0.7$ GeV. For $k_\gamma$ we take the values of the $J/\psi(1S)$ and $\eta_c(1S)$ 
masses from the PDG \cite{Beringer:1900zz}. In table \ref{tab:Gammac1S}, we show the size of the different contributions to $\Gamma_{J/\psi (1S) \rightarrow \eta_c(1S)\gamma}$. 
The ${\cal O}(\als)$ and ${\cal O}(\als^2)$ corrections are evaluated at the mass scale. 
The ${\cal O}(\als(m))$ corrections in the RS' and on-shell scheme are equal. 
Renormalon effects first appear at ${\cal O}(\als^2(m))$ and make this expansion more convergent. 
Yet, as we have taken a small value of $\nu_f$, the ${\cal O}(\als^2(m))$ term is still large. There are no ${\cal O}(v)$ corrections. 
The ${\cal O}(v^2)$ correction can be evaluated 
at different orders in $N$, and for different  values of the factorization scale. One can easily deduce its size by multiplying 
Fig. \ref{fig:charmn1}.b by -5/3 times the LO result. 
The value quoted in Table \ref{tab:Gammac1S} for the ${\cal O}(v^2)$ correction has been obtained for $N=3$ and $\nu=1.5$ GeV.
Unlike in Sec. \ref{sec:botn11}, in this case there are no scales of minimal sensitivity. The use of a finite $\nu_r$ significantly diminishes the factorization scale dependence of the result. Yet, we
also observe that a large scale dependence remains for small scales. Therefore, the value we take and quoted in Table \ref{tab:Gammac1S} for the 
${\cal O}(v^2)$ correction corresponds to $N=3$ and $\nu=1.5$ GeV, as we feel that smaller values of $\nu$ may yield unrealistic results. 

\begin{table}[hb]
\ra{1.1}
\begin{center}
$$
\begin{array}{|l|c|c|c|c|c|c|}
\hline
   &{\rm LO}&{\cal O}(\als)&{\cal O}(\als^2)&{\cal O}(v^2)&\als \times {\cal O}(\als^2)&v\times{\cal O}(v^2)\\
\hline
\delta \Gamma \;({\rm keV}) &2.34 &  0.33 & 0.16 & -0.71 & 0.05 & -0.30\\
\hline
\bottomrule
\end{array}
$$ 
\end{center}
\caption{\label{tab:Gammac1S} \it 
The leading and subleading contributions to $\Gamma_{J/\psi (1S) \rightarrow \eta_c(1S)\gamma}$. 
The last two numbers are error estimates obtained by multiplying the subleading ${\cal O}(\als^2)$ contribution by $\als$ 
and the subleading ${\cal O}(v^2)$ contribution by $v$.}
\end{table}

\begin{figure}
\includegraphics[width=0.7\textwidth]{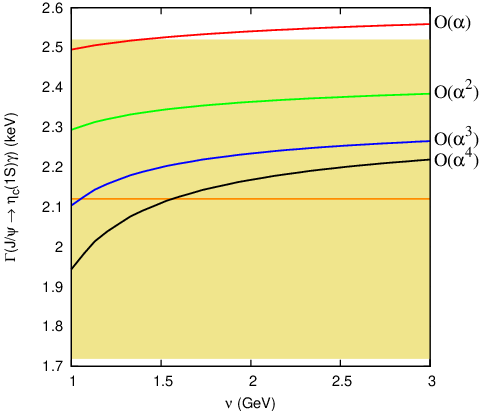}
\caption{\label{fig:charmn1decay} \it Plot of $\Gamma_{J/\psi(1S) \rightarrow \eta_{c}(1S)\gamma}$ using the static potential $V_{\RS'}^{(N)}$ at different orders in perturbation theory: $N=0,1,2,3$ with $\nu_r=\nu_f=0.7$ GeV.
The horizontal line is our central value and the yellow band our final error estimate. 
}
\end{figure}

If we sum all the contributions of Table \ref{tab:Gammac1S} we obtain 2.12 keV, which is quite close to the LO 2.34 keV value. This is due to the strong cancellation between the
 $\als$ and $v$ corrections. The main source of uncertainty comes from higher order terms. Because of the strong cancellation between the
 $\als$ and $v$ terms, we feel that adding a power of 
 $v$ to the overall correction would underestimate the error. Instead, we take the $v\times{\cal O}(v^2)$ contribution 
 in Table \ref{tab:Gammac1S} as our estimate of the subleading correction, as it is the biggest possible individual term. This term alone produces an error of order 15\%.  
This error is much bigger than the error one would obtain only considering scale variations (see Fig. \ref{fig:charmn1decay}), or if we do the evaluation with $N=2$ instead of $N=3$ (see, again, Fig. \ref{fig:charmn1decay}). It is also bigger than the error associated with variations of $\nu_f$. All these errors are associated with higher order effects. 
 We do not include those, in order to avoid double counting. The only other source of theoretical error that we include is the one due to $N_m$ (for the evaluation of this error we also take into account the correlation with the charm mass value). Besides the theoretical error, we also include the error associated with the QCD parameters, even though its size is typically smaller than the theoretical error. For $\als$ we take the variation $\als(M_z)=0.118\pm0.001$ \cite{Beringer:1900zz}. For the variation of the $\MS$ charm mass we take $m_c(m_c)=1.25\pm 0.04$ GeV (see, for instance, \cite{Signer:2008da}). In summary, we obtain the following result for the different error contributions:
\be
\Gamma_{J/\psi(1S) \rightarrow \eta_{c}(1S)\gamma}
=2.12\pm 0.30({\cal O}(v^3)){}^{+0.21}_{-0.23}(N_m){}^{-0.02}_{+0.02}(\als){}^{-0.10}_{+0.11}(m_{\MS}) 
\;{\rm keV}
\,,
\ee
%0.381142(theory)
which, after combining the errors in quadrature, reads
 \be
\Gamma_{J/\psi(1S) \rightarrow \eta_{c}(1S)\gamma}
=2.12(40)\;{\rm keV}
\,.
%0.396696
\ee
\begin{figure}
\includegraphics[width=0.8\textwidth]{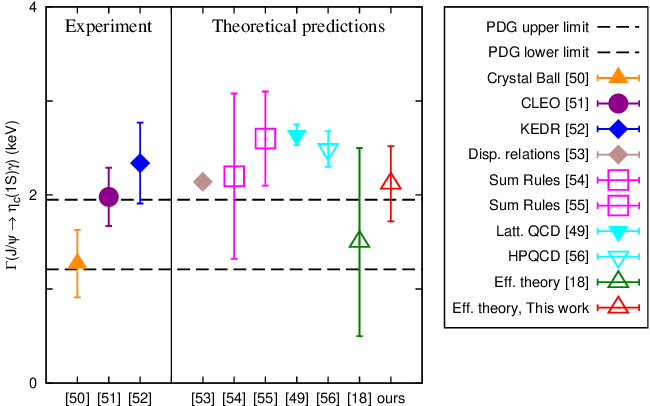}
\caption{\label{fig:charm_comparative} \it Comparison of different theoretical and experimental predictions for $\Gamma_{J/\psi \rightarrow \eta_{c}\gamma}$. 
}
\end{figure}
This corresponds to a branching fraction of $2.28\times 10^{-2}$.

We can now compare this with previous determinations of this decay. As in Ref.~\cite{Becirevic:2013gy}, we summarize the 
comparison in Fig. \ref{fig:charm_comparative}. Unlike in that reference, we do not include the values obtained in Refs.~\cite{Voloshin:2007dx,Eichten:2007qx}
assigned to potential models. 
They correspond to the LO computation in our notation (see Table \ref{tab:Gammac1S}). The difference with our value is (mainly) due to the different value of the 
charm mass. Note that in our case the charm mass is not a free parameter; rather, it is fixed by the value of the $\MS$ mass. We could still vary the 
RS' mass by changing $\nu_f$, but this effect would be compensated by the ${\cal O}(\als,v)$ effects. We now compare with the 
EFT computation of Ref. \cite{Brambilla:2005zw}. It is equivalent to ours, setting $N=0$, $\nu_r=\infty$, and eliminating the 
${\cal O}(\als^2)$ corrections. When we do so, we can get agreement with their number if we, as they do, set a very low value for the factorization scale (there are minor differences coming from the values of the heavy quark masses used). We see a very strong scale dependence in this regime. In this paper we restrict ourselves to values of $\nu$ where we get stable results after the resummation of the soft logarithms. This produces much bigger numbers, which, however, get reduced by increasing $N$. 
Either way, it should be emphasized that both results are perfectly compatible within errors. We also refer to 
Fig.~\ref{fig:charm_comparative} for comparison with the existing experimental numbers \cite{Gaiser:1985ix,Mitchell:2008aa,Anashin:2010nr},  and other theoretical predictions using either dispersion relations/sum rules 
\cite{Shifman:1979nx,Khodjamirian:1983gd,Beilin:1984pf} or lattice simulations \cite{Donald:2012ga,Becirevic:2013gy}. Our result is basically compatible with all of them within errors. We can discriminate very low values of the decay and start to have tensions with the Crystal Ball determination. 

\subsection{$P$-wave decays}

\begin{table}[hb]
\ra{1.1}
\begin{center}
$$
\begin{array}{|l|c|c|c|c|c|c|}
\hline
   &{\rm LO}&{\cal O}(\als)&{\cal O}(\als^2)&{\cal O}(v^2)&\als \times {\cal O}(\als^2)&v\times{\cal O}(v^2)\\
\hline
\delta \Gamma_{h_b(1P) \rightarrow \chi_{b0}(1P)\gamma} \;({\rm eV}) & 0.895  & 0.078  & 0.044 & -0.054 & 0.009 & -0.013\\
\hline
\delta \Gamma_{h_b(1P) \rightarrow \chi_{b1}(1P)\gamma} \;({\rm eV}\times 10^{-3}) & 8.86  & 0.77  & 0.43 & -1.08 &  0.09& -0.27\\
\hline
\delta \Gamma_{\chi_{b2}(1P) \rightarrow h_b(1P)\gamma} \;({\rm eV}) &0.113 & 0.010  & 0.006 & -0.011 & 0.001 & -0.003\\
\hline
\bottomrule
\end{array}
$$ 
\end{center}
\caption{\label{tab:Gammab2P} \it The leading and subleading contributions to $\Gamma_{h_b(1P) \rightarrow \chi_{b0}(1P)\gamma}$, $\Gamma_{h_b(1P) \rightarrow \chi_{b1}(1P)\gamma}$ and 
$\Gamma_{\chi_{b2}(1P) \rightarrow h_b(1P)\gamma}$, respectively. 
The last two numbers are error estimates obtained by multiplying the subleading ${\cal O}(\als^2)$ contribution by $\als$ 
and the subleading ${\cal O}(v^2)$ contribution by $v$.}
\end{table}

We now compute the $P$-wave decays for $n=2$ bottomonium (though they could end up being of academic interest because of the very small energy differences). 
In this case we have several decays (see Eq. (\ref{nP=nP})). The differences among them are spin factors, which weight the ${\vec p}^2$ matrix element  differently (there are also important differences for $k_{\gamma}$ depending on the decay). 
From the physical point of view 
the situation is similar to the two previous sections, as the squared wave function still has a single maximum, though more weighted 
at somewhat smaller scales. 
Our central values for the decays 
$\Gamma_{h_b(1P) \rightarrow \chi_{b0}(1P)\gamma}$, $\Gamma_{h_b(1P) \rightarrow \chi_{b1}(1P)\gamma}$, and 
$\Gamma_{\chi_{b2}(1P) \rightarrow h_b(1P)\gamma}$ are obtained using Eq. (\ref{nP=nP}) with $N=3$, $\nu=1.5$ GeV, and $\nu_f=\nu_r=0.7$ GeV. They are compatible with the numbers obtained in \cite{Brambilla:2005zw} if we account for the different $k_{\gamma}$ and a trivial misprint (keV $\rightarrow$ eV). For $k_\gamma$ we take the masses of the different $P$-wave states from the PDG \cite{Beringer:1900zz}. 
In table \ref{tab:Gammab2P} we show the size of the different contributions. The ${\cal O}(\als)$ and ${\cal O}(\als^2)$ corrections are evaluated at the mass scale. 
The ${\cal O}(\als(m))$ corrections are equal in the RS' and on-shell scheme. Renormalon effects first appear at 
${\cal O}(\als^2(m))$ and make this expansion more convergent, yet, as we have taken a small value of $\nu_f$, the ${\cal O}(\als^2(m))$ term is still relatively large. There are no ${\cal O}(v)$ corrections. 
The ${\cal O}(v^2)$ correction can be evaluated 
at different orders in $N$ and for different  values of the factorization scale. One can easily deduce its size by multiplying Fig. 
\ref{fig:botn21}.b by -5/3 times the LO result. 
Unlike in Sec. \ref{sec:botn11}, in this case there are no scales of minimal sensitivity. The use of a finite $\nu_r$ significantly diminishes the factorization scale dependence of the result. Yet, we observe that a strong scale dependence remains for small scales. Therefore, the value we  quote in Table \ref{tab:Gammab2P} for the 
${\cal O}(v^2)$ correction corresponds to $N=3$ and $\nu=1.5$ GeV, as we feel that smaller values of $\nu$ may yield unrealistic results. 
Note that, unlike the ${\cal O}(\als)$ corrections, this contribution is weighted differently for each decay. 
Therefore, properly weighted differences of these decays may yield absolute determinations of the $\langle {\vec p}^2 \rangle_{21}$ 
matrix element. In any case, the relative sign between 
the ${\cal O}(v^2)$ and ${\cal O}(\als(m))$ corrections produces cancellations between these terms. The magnitude of this cancellation 
depends on the specific decay mode, but it is large in all cases. 

\begin{figure}
\includegraphics[width=0.45\textwidth]{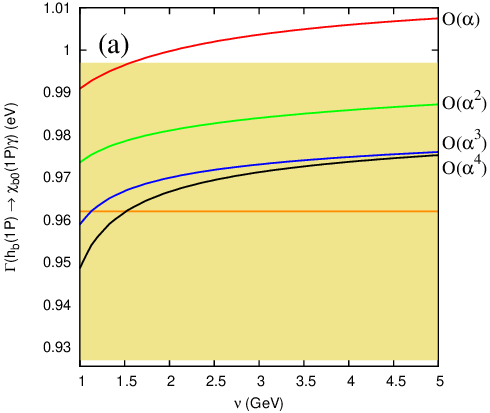}
\includegraphics[width=0.45\textwidth]{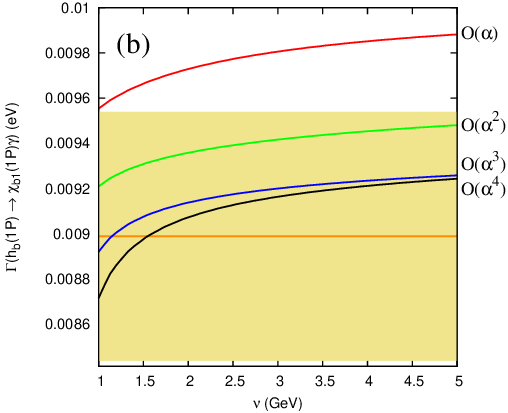}
\includegraphics[width=0.45\textwidth]{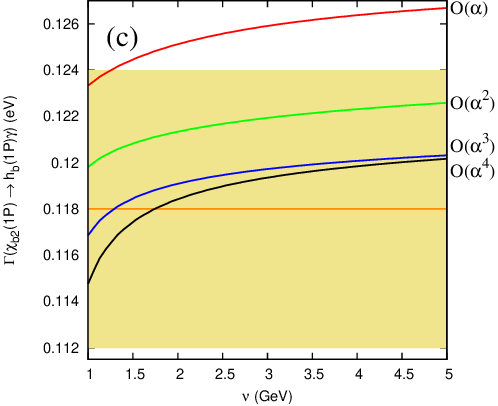}
\caption{\label{fig:botPdecay} \it Plot of $\Gamma_{h_b(1P) \rightarrow \chi_{b0}(1P)\gamma}$, $\Gamma_{h_b(1P) \rightarrow \chi_{b1}(1P)\gamma}$ and $\Gamma_{\chi_{b2}(1P) \rightarrow h_b(1P)\gamma}$ using the static potential $V_{\RS'}^{(N)}$ at different orders in perturbation theory: $N=0,1,2,3$ with $\nu_r=\nu_f=0.7$ GeV.
The horizontal line is our central value and the yellow band our final error estimate. 
}
\end{figure}

In order to estimate the errors we proceed analogously to the two previous sections. 
We take the $v\times{\cal O}(v^2)$ contribution of Table \ref{tab:Gammab2P} as our estimate of the subleading correction, as it is the biggest possible individual term. 
The only other source of theoretical error that we include is the one due to $N_m$ (for the evaluation of this error we also take into account the correlation with the bottom mass value). Besides the theoretical error, we also include the error associated with the QCD parameters, even though its size is typically smaller than the theoretical error. For $\als$ we take the variation $\als(M_z)=0.118\pm0.001$ \cite{Beringer:1900zz}. For the variation of the $\MS$ bottom mass we take $m_b(m_b)=4.19\pm 0.03$ GeV. In summary, we obtain the following result for the different error contributions for the three decays: 
\bea
\Gamma_{h_b(1P) \rightarrow \chi_{b0}(1P)\gamma}
&=&0.962
\pm 0.013({\cal O}(v^3)){}^{+0.029}_{-0.002}(N_m){}^{-0.001}_{+0.001}(\als){}^{-0.013}_{+0.013}(m_{\MS}) 
\;{\rm eV}
\,,
\\
\Gamma_{h_b(1P) \rightarrow \chi_{b1}(1P)\gamma}
&=&8.99
\pm 0.27({\cal O}(v^3)){}^{+0.46}_{+0.07}(N_m){}^{-0.04}_{+0.04}(\als){}^{-0.12}_{+0.12}(m_{\MS}) 
\times 10^{-3}\;{\rm eV}
\,,
\\
\Gamma_{\chi_{b2}(1P) \rightarrow h_b(1P)\gamma}
&=&0.118
\pm 0.003({\cal O}(v^3)){}^{+0.005}_{+0.000}(N_m){}^{-0.000}_{+0.000}(\als){}^{-0.002}_{+0.002}(m_{\MS}) 
\;{\rm eV}
\,,
\eea
which, after combining the errors in quadrature, read
\bea
\Gamma_{h_b(1P) \rightarrow \chi_{b0}(1P)\gamma}
&=&0.962(35)\;{\rm eV}
\,,
\\
\Gamma_{h_b(1P) \rightarrow \chi_{b1}(1P)\gamma}
&=&8.99(55)\times 10^{-3}\;{\rm eV}
\,,
\\
\Gamma_{\chi_{b2}(1P) \rightarrow h_b(1P)\gamma}
&=&0.118(6)\;{\rm eV}
\,.
\eea
The errors are heavily dominated by theory.
They are much bigger than the error one would obtain only considering scale variations (see Fig. \ref{fig:botPdecay}), 
or if we do the evaluation with $N=2$ instead of $N=3$ (see, again, Fig. \ref{fig:botPdecay}). They are also bigger than the error associated with variations of $\nu_f$. 

\subsection{$\Upsilon(2S) \rightarrow \eta_b(2S)\gamma$}

We now compute the $\Upsilon(2S) \rightarrow \eta_b(2S)\gamma$ decay. We use Eq. (\ref{nS=nS}) with $n=2$, 
which depends on $\langle {p}^2 \rangle_{20}$. We observed in Fig. \ref{fig:botn20} that this object was not convergent in $N$. 
Therefore, the results of this section should be taken with some caution. 
\begin{table}[h]
\ra{1.1}
\begin{center}
$$
\begin{array}{|l|c|c|c|c|}
\hline
   &{\rm LO}&{\cal O}(\als)&{\cal O}(\als^2)&{\cal O}(v^2)\\
\hline
\delta \Gamma \;({\rm eV})& 0.640 & 0.056 & 0.031 & -0.059\\
\hline
\bottomrule
\end{array}
$$ 
\end{center}
\caption{\label{tab:Gammab2S} 
\it 
The leading and subleading contributions to $\Gamma_{\Upsilon(2S) \rightarrow \eta_b(2S)\gamma}$.}
\end{table}

\begin{figure}
\includegraphics[width=0.7\textwidth]{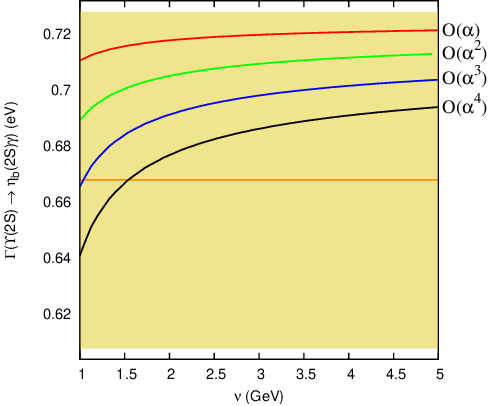}
\caption{\label{fig:botn20decay} \it Plot of $\Gamma_{\Upsilon(2S) \rightarrow \eta_b(2S)\gamma}$ for the case of the 
bottomonium ground state using the static potential $V_{\RS'}^{(N)}$ at different orders in perturbation theory: $N=0,1,2,3$ with $\nu_r=\nu_f=0.7$ GeV.
The horizontal line is our central value and the yellow band our final error estimate. 
}
\end{figure}

Our central value will be obtained by using Eq. (\ref{nS=nS}) with $N=3$, $\nu=1.5$ GeV, and $\nu_f=\nu_r=0.7$ GeV. For $k_\gamma$ we take the value of the $\Upsilon(2S)$ mass from the PDG \cite{Beringer:1900zz}. For the mass of the $\eta_b(2S)$ we take the recent value obtained by Belle \cite{Mizuk:2012pb}
for definiteness. Nevertheless, we should remark that a different value is obtained by using CLEO data \cite{Dobbs:2012zn} (if so our numbers can be trivially rescaled accordingly). In table \ref{tab:Gammab2S} we show the size of the different contributions to $\Gamma_{\Upsilon(2S) \rightarrow \eta_b(2S)\gamma}$. 
The ${\cal O}(\als)$ and ${\cal O}(\als^2)$ corrections are evaluated at the mass scale. 
The ${\cal O}(\als(m))$ corrections are equal in the RS' and on-shell scheme. Renormalon corrections first appear at 
${\cal O}(\als^2(m))$ and make this expansion more convergent, yet, as we have taken a small value of $\nu_f$, the ${\cal O}(\als^2(m))$ term is still large. There are no ${\cal O}(v)$ corrections. 
The ${\cal O}(v^2)$ correction can be evaluated 
at different orders in $N$ and for different values of the factorization scale. 
One can easily deduce its size by multiplying 
Fig. \ref{fig:botn20}.b by -5/3 times the LO result. The value quoted in Table \ref{tab:Gammab2S} for the 
${\cal O}(v^2)$ correction has been obtained for $N=3$ and $\nu=1.5$ GeV. For the $2S$ bottomonium state the use of a finite $\nu_r$, in particular $\nu_r=0.7$, does not significantly improve the $\nu_r=\infty$ computation. The factorization scale dependence is still significant and the convergence bad. Therefore, conservatively we take the ${\cal O}(v^2)$ term 
 as our estimate of the error associated with higher order corrections. This term alone produces an error of order 10\%. For the rest of the errors we proceed as in the previous sections. 
Overall, we obtain 
\be
\Gamma_{\Upsilon(2S) \rightarrow \eta_b(2S)\gamma}
=0.668
\pm 0.059({\cal O}(v^2)){}^{+0.004}_{-0.006}(N_m){}^{-0.002}_{+0.002}(\als){}^{-0.009}_{+0.009}(m_{\MS}) 
\;{\rm eV}
\,,
\ee
which after combining the errors in quadrature reads
\be
\label{th:2S2S}
\Gamma_{\Upsilon(2S) \rightarrow \eta_b(2S)\gamma}
=0.668(60)
\;{\rm eV}
\,.
\ee
In Fig. \ref{fig:botn20decay} we compare this result with the scale variation of the evaluation of the decay for different values of $N$. 
Note that our error is much bigger than the one from the factorization scale dependence, or from the difference between 
different $N$ evaluations. We believe an error analysis only based on any of those would underestimate the error. 

\subsection{$2S \rightarrow 1S\gamma$ decays}

The experimental situation of the $2S \rightarrow 1S$ radiative transitions has improved significantly over the last years. Whereas for the $2^{1}S_{0}\to 1^{3}S_{1}\gamma$ decay there are still no data available, this is not so for the 
$2^{3}S_{1}\to 1^{1}S_{0}\gamma$ decay, for which the PDG \cite{Beringer:1900zz} quotes the value $[3.9\pm1.5]\times 10^{-4}$ for the decay branching fraction. This translates into the following value for the decay:
\be
\Gamma^{(\rm exp)}_{\Upsilon(2S) \rightarrow \eta_b(1S)\gamma}
=12.5(4.9)\,
{\rm eV}
\,.
\ee 
This number comes from \cite{Aubert:2009as} BABAR (branching fraction $[3.9\pm1.1(stat)^{+1.1}_{-0.9}(syst)]\times 10^{-4}$) and updates the previous upper bound branching fraction $< 0.5 \times 10^{-3}$ ( or 
$\Gamma_{\Upsilon(2S) \rightarrow \eta_b(1S)\gamma}< 0.016$) produced by CLEOIII \cite{Artuso:2004fp}.

On the theoretical side, the $2S \rightarrow 1S$ radiative transitions are different from the previous transitions considered before. 
Now, we only know the leading nonvanishing order, see Eqs. (\ref{nVnot=nP}) and (\ref{nPnot=nV}), 
which scales as $\sim (k_{\gamma}^3/m^2)v^4$. It depends on the expectation values of ${\vec p}^2$, $\vec{r}^2$ and $V_{S^2}({\vec r})$ among different states ($n=1$ and $n=2$), which we have not studied so far (note that for those matrix elements we can only fix their relative sign but not the absolute one). Moreover, the $V_{S^2}({\vec r})/m^2$ potential is modulated by the Wilson coefficient $D^{(2)}_{S^2,s}$, which resums the large logarithms associated with the heavy quark mass. 

\begin{figure}
\includegraphics[width=0.9\textwidth]{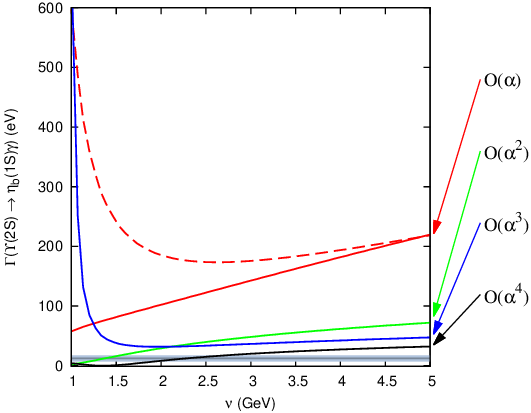}
\caption{\label{fig:botn2decay} \it Plot of $\Gamma_{\Upsilon(2S) \rightarrow \eta_b(1S)\gamma}$ using the static potential $V_{\RS'}^{(N)}$ at different orders in perturbation theory: $N=0,1,2,3$ with $\nu_f=0.7$ and
$\nu_r=\infty$. The dashed line corresponds to no resumming the hard logarithms: $D_{S^2,s}=\als(\nu)$. The blue band corresponds to the experimental value.
}
\end{figure}

The warning qualifications that we made in the previous section may also apply here, as the decay depends on the dynamics of the $2S$ bound state, for which we have observed problems of convergence in $N$ for $\langle {p}^2\rangle_{20}$. Nevertheless, it is worth repeating that the observables we are sensitive to now are different and, therefore, worth exploring. 
In Fig. \ref{fig:botn2decay} we show the theoretical predictions for $\Gamma_{\Upsilon(2S) \rightarrow \eta_b(1S)\gamma}$ after approximating the static potential at different orders in $N$ working at $\nu_r=\infty$ and $\nu_f=0.7$ GeV (see solid lines). In other words, we just add an $r$-independent constant to the static potential. We actually see a nicely convergent pattern for the decay, the magnitude of which decreases quite significantly as we increase $N$ (by around an order of magnitude) and approaches the experimental value. 

\begin{figure}
\epsfig{figure=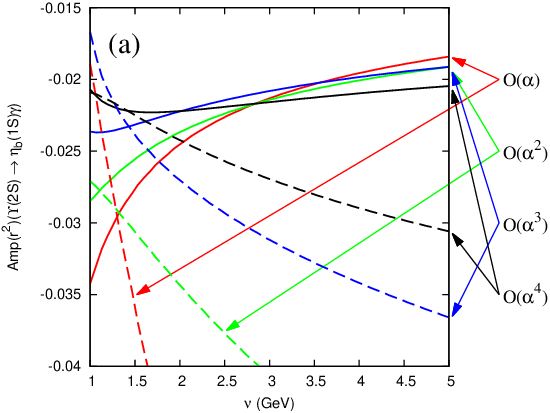,height=6.95cm,width=8.35cm}
\hspace{0.25cm}
\epsfig{figure=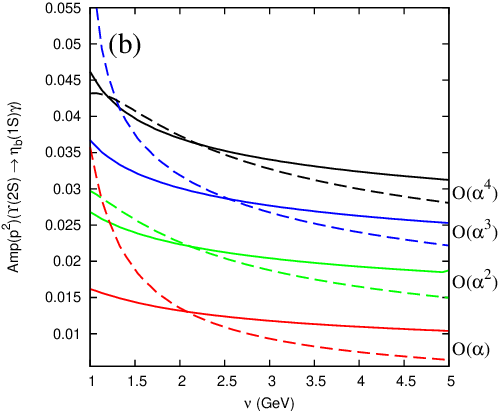,height=6.95cm,width=7.65cm} \\[3ex]
\epsfig{figure=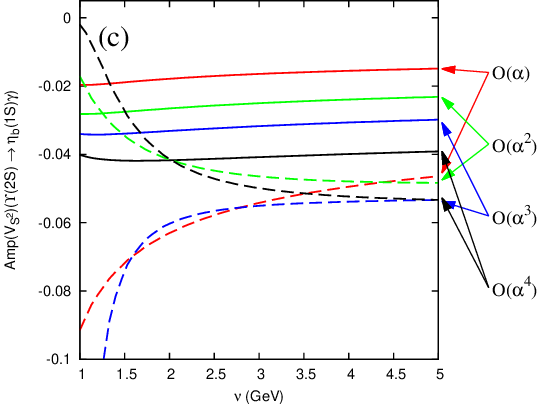,height=6.95cm,width=8.35cm}
\caption{\label{fig:bot2S1Smat_el} \it Plot of 1st, 2nd and 3rd term inside the brackets of Eq. (\ref{nVnot=nP}) 
using the static potential $V_{\RS'}^{(N)}$ at different orders in perturbation theory: $N=0,1,2,3$. The dashed lines correspond to the  
$\nu_f=0.7$ GeV and
$\nu_r=\infty$ case, and the continuous lines to the $\nu_f=\nu_r=0.7$ GeV case.
}
\end{figure}

In order to understand this result it is convenient to 
study the magnitude of the different terms that contribute to Eq. (\ref{nVnot=nP}). We display the terms inside the brackets of 
Eq. (\ref{nVnot=nP}) in Fig. \ref{fig:bot2S1Smat_el} with 
$\nu_r=\infty$ and $\nu_f=0.7$ GeV (see dashed lines). Note that they are ${\cal O}(v^2) \sim 0.06$, up to prefactors. 
We observe a very nice convergence pattern for the ${}_{10}\langle r^2\rangle_{20}$ associated term. The convergence of the ${}_{10}\langle p^2\rangle_{20}$ term is not as good, and even less for $ {}_{10}\langle V_{S^2}({\vec r})\rangle_{20}$ (for scales below 2.5 GeV). In any case, there is a very strong cancellation between the different terms in the decay. This makes the total sum of these terms smaller than the magnitude of each of them. The bulk of this effect is independent of the factorization scale and gets strongly magnified as we increase $N$ (see, again, the solid lines of Fig. \ref{fig:botn2decay}). Therefore, it does not seem to be a numerical accident for a specific $N$ or factorization scale.  

If we switch off the resummation of the hard logarithms and work at $N=0$ 
with $\nu_r=\infty$, our computation is equivalent to the analysis performed in Ref. \cite{Brambilla:2005zw}. In that reference a
very large value for the decay was obtained.  We show our equivalent computation as the dashed line in Fig. \ref{fig:botn2decay}. If 
we set $\nu=1$ GeV, the value used in that reference, we obtain $\Gamma_{\Upsilon(2S) \rightarrow \eta_b(1S)\gamma} \simeq 0.6$ keV
($\simeq 0.649$ keV if we use the mass $m_b=4730$ MeV used in that reference). Therefore, the introduction of the hard logarithms is crucial to make the decay transition width smaller for $N=0$ 
at small scales. As we increase $N$ this effect is less important, and the decay width gets small irrespective of the resummation of hard logarithms (yet, the final value may change by a factor 3 or 4, especially at small scales). At this stage we would like to emphasize that the computation of the 
decay shows a nicely convergent pattern in $N$, as we can see in Fig.~\ref{fig:botn2decay}, rapidly approaching the experimental number.

\begin{figure}
\includegraphics[width=1.05\textwidth]{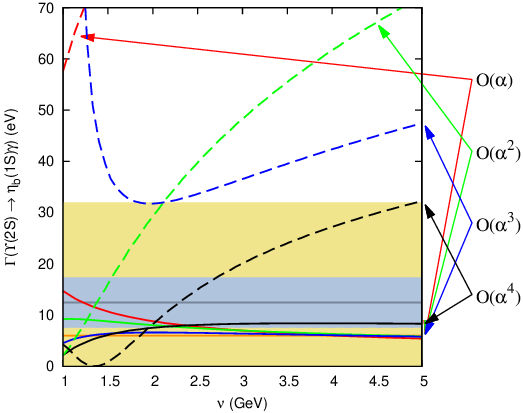}
\caption{\label{fig:botn2decaymur} \it 
Plot of $\Gamma_{\Upsilon(2S) \rightarrow \eta_b(1S)\gamma}$ using the static potential $V_{\RS'}^{(N)}$ at different orders in perturbation theory: $N=0,1,2,3$ with $\nu_f=\nu_r=0.7$ GeV (solid lines) and $\nu_f=0.7$ GeV, $\nu_r=\infty$ (dashed lines). The blue (darker) band corresponds to the experimental value and the yellow (lighter) band to our theoretical prediction.
}
\end{figure}

As in previous sections we can try to improve the previous results by exactly incorporating the correct asymptotic short distance behavior of the 
static potential in the solution of the Schroedinger equation. Typically, the convergence is accelerated and the factorization scale dependence greatly diminishes. We show the behavior of the different contributions to the decay in Fig. \ref{fig:bot2S1Smat_el} 
(see solid lines). There is a strong cancellation between the second and third term in Eq. (\ref{nVnot=nP}) (compare the solid lines of Fig. \ref{fig:bot2S1Smat_el}.b and Fig. 
\ref{fig:bot2S1Smat_el}.c), whereas the first term is almost constant (see the solid lines of Fig. \ref{fig:bot2S1Smat_el}.a).
 We show the result for the decay in Fig. \ref{fig:botn2decaymur} with $\nu_r=0.7$ GeV, where we also compare with the $\nu_r=\infty$ case, and experiment. Note how this figure corresponds to a zoom of Fig. \ref{fig:botn2decay}, as the scale dependence is much smaller, as well as the size of the corrections. For $N=0$ we still have some scale dependence, but it basically vanishes for $N>0$ and $\nu>2$ GeV. Actually, the results are very stable against scale variations (with a nice plateau for $\nu>2$ GeV) and to the value of $N$. In order to 
get these results the resummation of the hard logs plays a crucial role, especially at low $\nu$. We also emphasize that the final numbers compare quite favorably with experiment. This is by far nontrivial, as there has been more than one order of magnitude reduction with respect to the original numbers obtained with a pure Coulomb potential. 

\begin{table}[h]
\ra{1.1}
\begin{center}
$$
\begin{array}{|l|c|c|c|c|}
\hline
   {\rm Prefactor}  \;({\rm keV})&{\cal A}(r^2)&{\cal A}({\vec p}^2)&{\cal A}(V_{S^2})& \Gamma  \;({\rm eV}) \\
\hline
 10.3342 & 0.022 & 0.039 & -0.042&6.3\\
\hline
\bottomrule
\end{array}
$$ 
\end{center}
\caption{\label{tab:Gammab2S1S} 
\it 
The prefactor, the terms inside the brackets of 
Eq. (\ref{nVnot=nP}), and the total decay width $\Gamma_{\Upsilon(2S) \rightarrow \eta_b(1S)\gamma}$.}
\end{table}

Therefore, we dare to give a value for, and assign errors to, the decay width. In order to produce our final numbers we proceed analogously to the previous sections. In Table \ref{tab:Gammab2S1S} we give the prefactor, the different matrix elements for $N=3$, $\nu=1.5$ GeV and $\nu_r=\nu_f=0.7$ GeV, as well as the total decay width. In order to estimate the error associated with subleading effects in $v$, we multiply the biggest of the three $v^2$ contributions by $v$, instead of multiplying the sum of the three contributions by $v$, as we cannot guarantee that the cancellation between different terms takes place at higher orders. The structure of the error estimate would then be
%10.3343*0.04*1/4*2* Sqrt[0.0063/10.3343]
\be
\label{deltav3}
\delta \Gamma^{(v)}_{\Upsilon(2S) \rightarrow \eta_b(1S)\gamma} =B [(Av^2+\delta v^3)^2-(Av^2)^2]
\simeq B (Av^2)2\times\delta v^3\simeq 0.005\, {\rm keV},
\ee
where $A$ is a small number and $\delta v^3\sim v\times{\cal O}(v^2)$. 
We check the reliability of this error estimate by replacing the theoretical masses that appear in the third term in Eq. (\ref{nVnot=nP}) by the physical ones (as our result is very sensitive to this term). This effect is subleading in $v$ and produces a shift with respect to the central value of 
order $\delta \Gamma_{\Upsilon(2S) \rightarrow \eta_b(1S)\gamma} \simeq 0.006 \, {\rm keV}$. We take this number 
(which is quite similar to the number obtained in Eq. (\ref{deltav3})) as our estimate of the higher order uncertainties. We do not 
dwell further on the analysis of the higher order uncertainties, as our main error will come from a strong dependence on 
$N_m$\footnote{We could reduce the dependence on $N_m$ by increasing $\nu_f$ (and $\nu_r$). The price one would pay is a stronger dependence on $\nu$.}. We also compute the error associated with 
$\als$ and $m_b$. Summarizing all the errors we obtain
 \be
 \label{Gammath2S1Serrors}
\Gamma^{(\rm th)}_{\Upsilon(2S) \rightarrow \eta_b(1S)\gamma}
=0.006 
\pm 0.006({\cal O}(v^5)){}^{+0.026}_{-0.006}(N_m){}^{-0.001}_{+0.001}(\als){}^{-0.000}_{+0.000}(m_{\MS}) 
\, {\rm keV}.
\ee 
If we combine all the errors in quadrature our final number reads
\be
 \label{Gammath2S1Sfinal}
\Gamma^{(\rm th)}_{\Upsilon(2S) \rightarrow \eta_b(1S)\gamma}
=6^{+26}_{-06} \, {\rm eV}.
\ee 
The error is completely dominated by theory.  It completely covers the experimental prediction. Note that the same error is obtained using the scale variation of the $N=3$, $\nu_r=\infty$ result, which does not depend on $N_m$ but is typically less precise.

\begin{figure}
\includegraphics[width=0.75\textwidth]{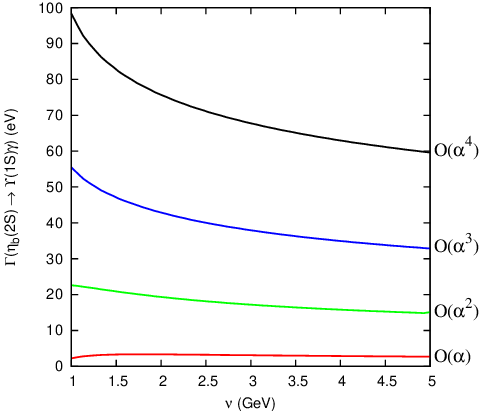}
\caption{\label{fig:botn2decayeta2} \it Plot of $\Gamma_{\eta_b(2S) \rightarrow \Upsilon(1S)\gamma}$ using the static potential $V_{\RS'}^{(N)}$ at different orders in perturbation theory: $N=0,1,2,3$ with $\nu_r=\nu_f=0.7$ GeV.
}
\end{figure}

Overall, we conclude that $\Gamma_{\Upsilon(2S) \rightarrow \eta_b(1S)\gamma}$ is relatively suppressed with respect its natural size by a 
very large cancellation between the $\langle p^2\rangle$ and $\langle V_{S^2}\rangle$ terms. This makes the total matrix element smaller\footnote{And an ideal place to measure $|{}_{10}\langle r^2 \rangle_{20}|$.}. The fact that it enters as $v^4$ magnifies this effect. A confirmation of this picture would come from the evaluation (and experimental determination) of $\Gamma_{\eta_b(2S) \rightarrow \Upsilon(1S)\gamma}$, which we expect to be much larger because the relative sign between these two terms changes. Actually, this is what we 
find, as one can see in Fig. \ref{fig:botn2decayeta2}. On the other hand, for this decay, there is no convergent pattern in $N$. Therefore, we do not dare to make any error analysis, and only estimate the decay to be around $\Gamma^{(\rm th)}_{\eta_b(2S) \rightarrow \Upsilon(1S)\gamma} \sim 80 \, {\rm eV}$.

We can compare Eq. (\ref{Gammath2S1Sfinal}) with the recent lattice simulation of Ref. \cite{Lewis:2012bh}. As our computation is ${\cal O}(v^4)$ we should compare with their  ${\cal O}(v^4)$ result. In matrix element units this corresponds to the number 0.080(5) in Table II of Ref. \cite{Lewis:2012bh} (the experimental number is 0.035(7)). Our central value corresponds to $0.025^{+0.031}_{-0.025}$. Nevertheless, a proper comparison would require the incorporation of the renormalization group improved Wilson coefficient, $D^{(2)}_{S^2,s}$, in the lattice analysis\footnote{This has not been done so far. We stress that this effect could be quite important. Actually one only has to incorporate the very same logs that we are incorporating here, as the leading logs are scheme independent.}. If we switch it off in our analysis our result gets strongly scale dependent and we can get agreement with their results 
for a scale of around 1 GeV. In this respect we cannot avoid mentioning that we expect some dependence on the lattice spacing of the NRQCD matrix elements, as, in general, it is not possible to obtain the continuum limit for them. In any case, we now face an interesting situation: In Ref. \cite{Lewis:2012bh} agreement with experiment was only obtained after the inclusion of the ${\cal O}(v^6)$ operators (again using tree-level Wilson coefficients). Note that this implies a complete breakdown of the $v$ expansion for bottomonium, as the ${\cal O}(v^6)$ correction would be as important as the ${\cal O}(v^4)$ term. 
On the other hand, our picture is different, and it is possible 
to obtain agreement with experiment with an ${\cal O}(v^4)$ 
computation (and the help of the renormalization group at small scales).

\section{Conclusions}
\label{sec:con}
We have computed the magnetic dipole transitions between low-lying heavy quarkonium states in a 
model-independent way. We have used the weak-coupling version of pNRQCD with the static 
potential exactly incorporated in the LO Hamiltonian. 
The precision we have reached is $k_{\gamma}^3/m^2\times{\cal O}(\als^2,v^2)$ and $k_{\gamma}^3/m^2\times{\cal O}(v^4)$ for the allowed and forbidden transitions, respectively. Large logarithms associated with the heavy quark mass scale have also been 
resummed. The effect of the new power counting was found to be large, and the exact treatment of the soft logarithms of the static potential made the factorization scale dependence much smaller. The convergence for the $b\bar b$ ground state was quite good, and also quite reasonable for the $c\bar c$ ground state and the $b\bar b$ $1P$ state. For all of them we have given solid predictions, which we summarize here:
\bea
\Gamma_{\Upsilon(1S) \rightarrow \eta_b(1S)\gamma}
&=&15.18(51) \;{\rm eV}
\,,
\\
\Gamma_{J/\psi(1S) \rightarrow \eta_{c}(1S)\gamma}
&=&2.12(40)\;{\rm keV}
\,,
\\
\Gamma_{h_b(1P) \rightarrow \chi_{b0}(1P)\gamma}
&=&0.962(35)\;{\rm eV}
\,,
\\
\Gamma_{h_b(1P) \rightarrow \chi_{b1}(1P)\gamma}
&=&8.99(55)\times 10^{-3}\;{\rm eV}
\,,
\\
\Gamma_{\chi_{b2}(1P) \rightarrow h_b(1P)\gamma}
&=&0.118(6)\;{\rm eV}
\,.
\eea

For the $2S$ decays the situation is less conclusive. The ${\cal O}(v^2)$ correction of the $\Upsilon(2S) \to \eta_b(2S)\,\gamma$ decay suffered from a bad convergence in $N$, producing relatively large errors for our prediction (see Eq. (\ref{th:2S2S})). Some of the ${\cal O}(v^2)$ matrix elements of the $\eta_b(2S) \to \Upsilon(1S)\,\gamma$ decay also suffered from this bad convergence. This made it impossible to give a reliable error estimate for this transition, as such terms correspond to the leading (and only known so far) order expression (moreover, they should be squared in the decay). The situation is completely different for the $\Upsilon(2S) \to \eta_b(1S)\,\gamma$ transition. The reason is that the problematic 
${\cal O}(v^2)$ matrix elements appear in a different combination for this decay, so that they cancel to a large extent. This led to a nicely convergent sequence in $N$, where the resummation of the hard logarithms played an important role. Our final figure was  
\be
 \Gamma^{(\rm th)}_{\Upsilon(2S) \rightarrow \eta_b(1S)\gamma}
=6^{+26}_{-06} \, {\rm eV}.
\ee 
This number is perfectly consistent with existing data, so that the previous disagreement with experiment for the $\Upsilon(2S) \to \eta_b(1S)\,\gamma$ decay fades away. 

The error of the above figures is dominated by theory, in most cases by the lack of knowledge of higher order effects. The determination of the origin and nature of those effects may significantly diminish the errors. Typically, they may come from loop effects, so it may happen that they 
effectively count as 
${\cal O}(\als v^2)$, implying smaller errors. In any case, let us note that the static potential becomes steeper as we increase $N$. 
Therefore, the transfer energy between the heavy quarks is bigger and the effective alpha and radius of the bound state become smaller than 
what one would deduce from a pure LO Coulomb evaluation. This means that the weak-coupling approximation works better than one would expect {\it a priori} for those systems. This is good news for weak-coupling analysis of the properties of the lowest-lying heavy quarkonium resonances.

We have not incorporated the error associated with $k_{\gamma}$ in our final numbers. Therefore, strictly speaking, our figures are theoretical predictions of $\Gamma/k_{\gamma}^3$.  
In some cases the associated error would be small. Yet, we have chosen to work in this way since, for some decays, the experimental value of $k_{\gamma}$ 
is still uncertain. It is trivial for the reader to introduce such error.  

We have also computed some expectation values like the electromagnetic radius, $\langle r^2 \rangle$, or $
\langle { p}^2 \rangle$. We find $\langle r^2 \rangle$ to be nicely convergent in all cases, whereas the convergence of 
$\langle { p}^2 \rangle$ is typically worse. We have found that $\langle p^2 \rangle$ is more or less constant with $n$, and 
$\sqrt{\langle r^2 \rangle}$ is more or less linear with $n$. This is the same behavior one finds with a logarithmic potential. Since the early days of heavy quarkonium it is well known from potential models that such a potential effectively describes the spectrum of the bottomonium and charmonium systems \cite{Quigg:1979vr}. We find it rewarding that the QCD potential can simulate such behavior after the inclusion of the 
running of $\als$.

The computation of $\langle r^2 \rangle$ and $\langle { p}^2 \rangle$ (and the binding energies) also yields a very nice confirmation of the renormalon dominance picture. This predicts in the on-shell scheme that, on the one hand, the binding energy should diverge with $N$ but, on the other, $\langle r^2 \rangle$ and $\langle { p}^2 \rangle$ should produce convergent sequences in $N$. We have observed this effect in full glory in our analysis. 

\acknowledgments{We gratefully acknowledge several clarifications remarks from A. Vairo on some aspects of Ref. \cite{Brambilla:2005zw}.
This work was partially supported by the spanish 
grants FPA2010-16963, FPA2010-21750-C02-02 and FPA2011-25948, by the catalan grant SGR2009-00894, 
by the European Community-Research
Infrastructure Integrating Activity 'Study of Strongly Interacting Matter'
(HadronPhysics3 Grant No. 283286), by the Spanish Ingenio-Consolider 2010
Program CPAN (CSD2007-00042) and also by the U.S. Department of Energy,
Office of Nuclear Physics, under contract DE-AC02-06CH11357.
}

%\begin{references}

%\end{references}

\end{document}